\documentclass[english]{article}

\usepackage{amsthm}
\usepackage{amsmath}
\usepackage{amssymb}
\usepackage{youngtab}

\usepackage{jheppub}

\newcommand{\tp}{\tilde{p}}

\newcommand{\ZZ}{{\mathbb Z}}
\newcommand{\RR}{{\mathbb R}}
\newcommand{\CC}{{\mathbb C}}
\newcommand{\ra}{\rightarrow}
\newcommand{\cN}{{\mathcal N}}
\newcommand{\Tr}{{\rm Tr}}
\newcommand{\bQ}{{\bar Q}}
\newcommand{\eps}{{\epsilon}}
\newcommand{\beps}{{\bar\epsilon}}
\newcommand{\id}{{\mathbf 1}}
\newcommand{\sign}{{\rm sign}}
\newcommand{\tQ}{{\tilde Q}}
\newcommand{\kb}{{k_b}}

\author[a]{Anton Kapustin,} 
\author[b]{Brian Willett} 
\affiliation[a]{Department of Physics, California Institute of Technology,\\ 1200 E. California Blvd, Pasadena, CA} 
\affiliation[b]{Institute for Advanced Study, Einstein Drive, Princeton, NJ} 
\emailAdd{kapustin@theory.caltech.edu} 
\emailAdd{bwillett@ias.edu} 

\begin{document}

\title{Wilson loops in supersymmetric Chern-Simons-matter theories and duality}

\author{\abstract{We study the algebra of BPS Wilson loops in 3d gauge theories with $\cN=2$ supersymmetry and Chern-Simons terms. We argue that new relations appear on the quantum level, and that in many cases this makes the algebra finite-dimensional. We use our results to propose the mapping of Wilson loops under Seiberg-like dualities and verify that the proposed map agrees with the exact results for expectation values of circular Wilson loops. In some cases we also relate the algebra of Wilson loops to the equivariant quantum K-ring of certain quasi projective varieties. This generalizes the connection between the Verlinde algebra and the quantum cohomology of the Grassmannian found by Witten.} }

\maketitle

\section{Introduction}

Despite impressive evidence for Seiberg duality for $\cN=1$ $d=4$ supersymmetric gauge theories \cite{Seiberg:1994pq}, its physical origin remains somewhat mysterious. To clarify it, it would be helpful to understand how Seiberg duality acts on Wilson loop operators. In this paper we attack a different but related problem: we determine how Wilson loops map under the Giveon-Kutasov (GK) duality which relates $\cN=2$ $d=3$ Chern-Simons-matter theories. Just like the 4d Seiberg duality, the GK duality exists for all classical gauge groups. In the unitary case, the ``electric'' theory has gauge group $U(N_c)$ at Chern-Simons level $k$ and $N_f$ fundamental flavors, while the ``magnetic'' theory has gauge group $U(|k|+N_f-N_c)_{-k}$ and $N_f$ fundamental flavors (as well as some singlets). Clearly, this looks very similar to the 4d Seiberg duality. But this is more than mere similarity. The GK duality can be regarded as a consequence of another Seiberg-like duality proposed by Aharony \cite{Aharony:1997gp} which relates $\cN=2$ $d=3$ theories without Chern-Simons terms. In turn, the Aharony duality is related to 4d Seiberg duality via compactification on a circle \cite{Aharony:1997gp}. We hope that understanding the mapping of Wilson loops under the GK duality will help to determine the corresponding map for the Aharony duality, and then perhaps for the 4d Seiberg duality.

The GK duality has several simplifying features compared to the 4d Seiberg and Aharony dualities. First, by adding a suitable superpotential, one can deform it into a duality for $\cN=3$ $d=3$ theories which are superconformal not only on the classical level, but quantum-mechanically. We will argue that properties of supersymmetric Wilson loops are not affected by this deformation. Second, in the special case $N_f=0$ the GK duality reduces to the familiar level-rank duality of  Chern-Simons theories without matter, and one can hope that the intuition gained from studying the level-rank duality will be helpful. Third, in the case when the gauge group is unitary, one can relate the GK duality to dualities for 2d topological field theories.  In the case $N_f=0$ this was recognized a long time ago by Witten \cite{Witten:1993xi}. 

A crucial point is that BPS Wilson loops in $\cN=2$ $d=3$ gauge theories form a commutative algebra. Our approach is based on determining the algebra and then observing that there is a natural isomorphism between ``electric'' and ``magnetic'' algebras of Wilson loops. Then we show that normalized expectation values of Wilson loops related by this isomorphism coincide. This provides strong evidence that the isomorphism is indeed the duality map. 

To give some flavor of our results, we consider the example of a $U(2)$ theory with a Chern-Simons term at level $k$ and $N_f$ fundamental hypermultiplets, and let us analyze the cases with $k+N_f=4$.  In the case $k=4,N_f=0$, this is a pure Chern-Simons theory, and the Wilson loop representations are truncated so that the corresponding Young diagrams fit in a box of size $2 \times 2$. Wilson loops whose Young diagrams violate the constraint are not independent but expressed through those whose Young diagrams do satisfy the constraint. In other words, there are additional quantum relations in the Wilson loop algebra that are not explained by classical representation theory. For example, in the present case we have a relation (labeling Wilson loop operators by the corresponding Young diagrams):

\[ {\tiny \yng(3)} = 0 \]
Level-rank duality asserts this theory is equivalent to a $U(2)$ Chern-Simons theory at level $-4$, and it is well known that the correct mapping of Wilson loop representations is to simply transpose the Young diagram. 

If we take $N_f=1, k=3$, then we will find below a simple generalization of this story, namely, the Wilson loops are still truncated to fit in a $2 \times 2$ box, but this time via a relation:

\[ {\tiny \yng(3)} = - \; {\tiny \yng(2)} + q \]
where $q=e^{2 \pi \zeta}$, with $\zeta$ the FI term of the theory.  This is GK dual to a theory of the same type, and the isomorphism sends, for example:

\[ {\tiny \yng(1)} \rightarrow - \; {\tiny \yng(1)} - 1, \;\;\;\; {\tiny \yng(1,1)} \rightarrow {\tiny \yng(2)} + {\tiny \yng(1)} \]
We see that this is similar to the level-rank duality rule of transposing the Young diagram, although with extra terms with fewer boxes appearing on the RHS.  Finally, if we take a theory with $k=N_f=2$, then we can give the two flavors opposite vector masses, say, $m$ and $-m$, and we find the relation:

\[ {\tiny \yng(3)} = - (r+r^{-1}) {\tiny \yng(2)} + (q-1) {\tiny \yng(1)}  + q (r+r^{-1})\]
where $r=e^{2 \pi m}$, and the duality map acts by:

\[ {\tiny \yng(1)} \rightarrow - \; {\tiny \yng(1)} - r-r^{-1}, \;\;\;\; {\tiny \yng(1,1)} \rightarrow {\tiny \yng(2)} + (r+r^{-1}) {\tiny \yng(1)} + 1-q\]

Our determination of the algebra of Wilson loops makes use of the exact formulas for their expectation values derived in \cite{Kapustin:2009kz}. Another possible approach is to compactify the 3d theory on a circle and argue that the algebra of Wilson loops is isomorphic to the twisted chiral ring of the 2d effective theory. For $N_f=0$ this approach relates the Verlinde algebra of $U(N_c)$ to the quantum cohomology of a Grassmannian, as in \cite{Witten:1993xi}. It turns out that for $N_f>0$ one obtains a relation between the algebra of Wilson loops and the equivariant quantum cohomology of a certain vector bundle over a Grassmannian. The latter has not been computed in general, as far as we know, so our computation of the Wilson loop algebra provides a prediction for it. GK duality also implies that equivariant quantum cohomology rings of two natural bundles over the same Grassmannian are isomorphic. 

A.K. would like to thank Kentaro Hori for a useful discussion which contributed to our understanding of section 7. The work of A.K. was supported in part by the DOE grant DE-FG02-92ER40701, and that of B.W. was supported by DOE grant DE-FG02-90ER40542.

\section{BPS Wilson loops}
\label{bpsw}

Our main object of interest is a Wilson loop operator in $\cN=2$ $d=3$ supersymmetric gauge theory. An ordinary Wilson loop is defined as
$$
\Tr_R {\rm Pexp}\left(i\int_\gamma A_\mu dx^\mu\right),
$$
where $\gamma$ is a closed loop in space-time and $R$ is a representation of the gauge group $G$. If one is dealing with a supersymmetric gauge theory, it is natural to require the loop operator to preserve some supersymmetry. An $\cN=2$ $d=3$ theory has spinor supercharges $Q_\alpha$, $\alpha=1,2$ and $\bQ_\alpha$ which act on the components of the vector multiplet $(A,\sigma,\lambda, D)$ as follows:
\begin{align*}
\delta A_\mu & =  -\frac{i}{2} (\beps \gamma_\mu\lambda-\bar\lambda\gamma_\mu\eps)\\
\delta \sigma & =   \frac12(\beps\lambda-\bar\lambda\eps) \\
\delta \lambda & =  \frac12\gamma^{\mu\nu}\eps F_{\mu\nu}-D \eps+i\gamma^\mu\eps D_\mu\sigma \\
\delta\bar\lambda & =  \frac12\gamma^{\mu\nu}\beps F_{\mu\nu}+D \beps-i\gamma^\mu\beps D_\mu\sigma \\
\delta D & =  -\frac{i}{2} \beps\gamma^\mu D_\mu\lambda-\frac{i}{2} D_\mu\bar\lambda \gamma^\mu\eps +\frac{i}{2}[\beps\lambda,\sigma]+\frac{i}{2}[\bar\lambda\eps,\sigma].
\end{align*}
Here $\eps$ and $\beps$ are covariantly constant spinors. The usual Wilson loop is not invariant under any SUSY transformations. But for particular $\gamma$ one can modify the Wilson loop to make it half-BPS. For example, consider a loop operator
$$
W_R = \Tr_R {\rm Pexp}\left(i\int_\gamma (A_\mu dx^\mu-i\sigma d\ell)\right),
$$
 in $\RR^2 \times S^1$ wrapping $S^1$ and localized at a point in $\RR^2$. If we denote by $x^3$ the coordinate along $S^1$, then this Wilson loop is invariant if we require
$$
(\gamma^3+1)\eps=0,\quad (\gamma^3-1)\beps=0.
$$
With the usual choice of Pauli matrices, this means that this Wilson loop is invariant with respect to $Q_1$ and $\bQ_2$.

Note that an $\cN=2$ $d=3$ theory on $\RR^2\times S^1$ can be regarded as an $\cN=(2,2)$ $d=2$ theory on $\RR^2$ (with an infinite number of 2d fields representing the Kaluza-Klein excitations). Local operators in this theory annihilated by $Q_1$ and $\bQ_2$ are called twisted chiral operators, and it is well-known that they form a ring (the twisted chiral ring).
 {}From the 3d viewpoint, all elements of the twisted chiral ring arise from loop operators wrapping the circle. On the other hand, the 3d chiral ring (i.e. local operators annihilated by $\bQ_1$ and $\bQ_2$) descends to the 2d chiral ring under compactification.

An important conclusion is that to every $\cN=2$ $d=3$ theory one can attach a commutative ring of BPS loop operators. BPS Wilson loops are examples of such operators, but there may be others. While there are no other candidate BPS loop operators on the classical level, on the quantum level one also needs to consider disorder loop operators, such as vortex loops \cite{Kapustin:2012iw,Drukker:2012sr}. On general grounds, it is not obvious if BPS Wilson loops form a sub-ring or not. Nevertheless, unless the semi-classical intuition is completely wrong, a product of two Wilson loops cannot contain a disorder loop operator, and therefore Wilson loops must form a sub-ring. This is also suggested by the matrix model computing the expectation values of circular Wilson loops on $S^3$ \cite{Kapustin:2009kz,Jafferis:2010un}. Such Wilson loops are related to straight Wilson loops on $\RR^3$ provided the theory is superconformal. The matrix model tells us that an insertion of two Wilson loops in representations $R_1$ and $R_2$ wrapping two large circles on $S^3$ is equivalent to an insertion of a single Wilson loop in representation $R_1\otimes R_2$ wrapping a large circle on $S^3$. 

On the classical level, the algebra of BPS Wilson loop is simply the representation ring of the gauge group $G$. The matrix model tells us that all classical relations remain true on the quantum level. However, it may be possible that new ``quantum'' relations arise . Below we will use the matrix model to propose in some cases what these new relations are and check that they are compatible with dualities. 

It is important to understand the dependence of the relations in the Wilson loop algebra on parameters such as real masses $m_a$ and the FI parameter $\zeta$. We will now argue that relations are always Laurent series in the parameters $r_a=\exp(2\pi m_a)$ and $q=\exp(2\pi\zeta)$.  Recall that $m_a$ and $\zeta$ can be thought of as expectation values of real scalars in background vector multiplets which couple to global symmetry currents (flavor and topological currents, respectively). Upon compactification on a circle of unit radius , these scalars are complexified, the imaginary parts coming from the expectation values of the background gauge fields along the compactified direction. $\cN=(2,2)$ $d=2$ supersymmetry requires BPS quantities such as relations in the BPS Wilson loop algebra to depend holomorphically on these complex scalars. But since the imaginary parts of these scalars are periodically identified with period $1$, BPS quantities must be Laurent series in the variables $r_a$ and $q$. In other words, the algebra of Wilson loops should be thought of as an algebra over the ring of Laurent series $\CC[[r_a,r_a^{-1},q,q^{-1}]]$. Duality maps must also be defined over this ring.

It was argued in  \cite{Kapustin:2006pk} that the algebra of loop operators in any TQFT must be defined over $\ZZ$. In this paper we are interested in 3d superconformal theories rather than TQFTs. Nevertheless, a suitable modification of the argument shows that in flat space-time the algebra of Wilson loops in a generic $\cN=2$ $d=3$ SCFT must be defined over the ring $\ZZ[r_a,r_a^{-1},q,q^{-1}]$. By a generic 3d SCFT we mean a 3d SCFT where the IR dimension of scalars in the vector multiplets is $1$. To see this, recall that by definition a loop operator is a trace of a line ``operator''. A line ``operator''  supported at a fixed point $p$ in space and extended in the time direction is not really an operator but a prescription to modify the theory at $p$. For example, in the case of ordinary line operators (meaning, not disorder line operators) this modification consists of tensoring the Hilbert space of the theory with a finite-dimensional vector space $R$ and modifying the Hamiltonian as follows:
$$
H\mapsto H\otimes \id_R+\Delta H_p,
$$
where $\Delta H_p$ depends only on the values of the fields at the point $p$. The evolution operator is then modified by a factor
$$
{\rm Pexp} \left(i\int dt\ \Delta H_p\right),
$$
where $t$ is the time coordinate.
The loop operator is the trace of this factor over $R$. (In the case of disorder line operators, one may also require the fields to have a prescribed singularity at $p$). A complex multiple of a BPS loop operator could then be understood as a modification of $\Delta H_p$ by an additive constant. But now we recall that BPS Wilson line ``operators'' in a generic SCFT are invariant under dilatations, and therefore the line ``operator'' which appears in their product must have the same property.  A shift of $\Delta H_p$ by a constant would violate dilatation-invariance, and therefore is impossible. Therefore a complex multiple of a BPS Wilson loop cannot appear in the OPE of BPS Wilson loops. One can nevertheless make sense of multiplication of a BPS Wilson loop by a positive integer $n$: one needs to replace $R$ with $R\otimes\CC^n$ and $\Delta H_p$ with $\Delta H_p\otimes \id_{\CC^n}$. In a supersymmetric theory it is natural to take $R$ to be a $\ZZ_2$-graded vector space and define the loop operator as a super trace of the evolution operator. Then the natural operation of the fermionic parity reversal multiplies the loop operator by $-1$. Combining these two natural operations, we can give a meaning to multiplication of a BPS Wilson loop operator by an arbitrary integer. Thus we expect that in flat space-time  there is a choice of generators in the algebra of BPS Wilson loops such that all relations have integral coefficients.

The above argument assumed that dilatation invariance is unbroken, and in particular the real masses and FI parameters vanish, but it can be easily generalized. The Hamiltonian $\Delta H_p$ may depend on the expectation values of bosons in background vector multiplets, and to leading order in spatial derivatives this dependence must be very simple thanks to supersymmetry and background gauge-invariance:
$$
\Delta H_p\mapsto \Delta H_p+\sum_a n_a (A_{a\, t}(p) - i\sigma_a(p)),
$$
where summation extends over all background vector multiplets, and $n_a$ are integers (eigenvalues of generators of the background gauge symmetries when acting on $R$). If we set the background gauge field $A_a$ to zero and the background scalar $\sigma_a$ to a constant $m_a$, then such a modification of the line ``operator'' multiplies the corresponding BPS Wilson loop wrapped on a circle of length $2\pi$ by a factor 
$$
\exp\left(2\pi \sum_a n_a m_a\right)=\prod_a r_a^{n_a}.
$$
Thus we expect that on $\RR^2\times S^1$ the coefficient ring of the algebra of BPS Wilson loops is the ring of Laurent series $\ZZ[[r_a,r_a^{-1},q,q^{-1}]]$. Moreover, since the structure constants of the algebra must be well-defined and integral if we specialize the variables $r_a$ and $q$ to $1$ (i.e. if we set the expectation values of background scalars to zero), the coefficient ring must in fact be the ring of Laurent polynomials $\ZZ[r_a,r_a^{-1},q,q^{-1}]$. 

In this paper we are actually computing the algebra of Wilson loops on $S^3$, not in flat space-time, and moreover localization methods force us to use nonstandard framing for Wilson loops. 
While conformal symmetry allows us to relate circular Wilson loops on $S^3$ and straight Wilson lines on $\RR^3$, switching to the standard framing introduces additional phases in the structure constants of the algebra. However, the dependence on the parameters $m_a$ and $\zeta$ is not affected by these framing phases, and we expect the coefficient ring on $S^3$ to be the ring of Laurent polynomials $\CC[r_a,r_a^{-1},q,q^{-1}].$ Note that this is a somewhat stronger result than what we obtained by compactification on a circle.

\section{Wilson loops in pure Chern-Simons theory}

Let us begin with $\cN=2$ $d=3$ super-Chern-Simons theory without matter. Such a theory is equivalent to pure Chern-Simons theory, since the scalars $\sigma$ and $D$ as well as the gaugino $\lambda$ are auxiliary. Moreover, the action for these fields is Gaussian:
$$
S_{aux}=\frac{ik}{4\pi} \int \Tr (2\sigma D-\bar\lambda\lambda),
$$ 
so we can use the equation of motion $\sigma=0$ to see that the BPS Wilson loop is equivalent to the ordinary Wilson loop. The algebra of Wilson loops in Chern-Simons theory is known as the Verlinde algebra. There are several interpretations of this algebra and accordingly several possible approaches to computing it:
\begin{itemize}
\item It is the $K^0$-ring of a modular tensor category (of conformal blocks in the corresponding WZW model);
\item It is the algebra of local operators in the gauged WZW model ($G/G$ model);
\item In the case $G=U(N_c)$, it is the quantum cohomology ring of the Grassmannian $Gr(N_c,\CC^{|k|})$ \cite{Witten:1988hf}.
\end{itemize}
All of these interpretations are two-dimensional in nature. For example, the second interpretation is based on the fact that Chern-Simons theory compactified on a circle is equivalent to the $G/G$-model, and Wilson loops wrapped on the compactification circle become local operators in the G/G model.   

We now explain an intrinsically 3d approach based on the $S^3$ matrix model. The advantage of this approach is that it can be relatively easily generalized to Chern-Simons-matter theories.\footnote{In section \ref{sec:Ktheory} we will see that the ``Grassmannian'' approach can also be generalized.}    This will involve determining certain ``quantum relations'' satisfied by the Wilson loop operators.  A quantum relation in the algebra of Wilson loops is a linear combination of Wilson loops (with integral coefficients) which vanishes as an operator in Hilbert space and therefore has zero expectation value on $S^3$. We therefore look for nontrivial vanishing of the matrix integral which computes the expectation values of circular Wilson loops on $S^3$. Each such vanishing gives a candidate quantum relation.

Some vanishings may be ``accidental', i.e. not coming from quantum relations.  One can try to weed them out in several different ways.  First, if a Wilson loop $W$ vanishes, then it should also be the case that any expression of the form $W W'$ vanishes in order that the relations define an ideal, and give rise to a consistent algebra.  A priori, there is no reason that the vanishing of the matrix integral computing the expectation value of the first expression would imply that of the second, so this is a non-trivial restriction.  Secondly, it has been argued above that quantum relations must involve only integer coefficients (perhaps after a suitable redefinition of generators). Another way to do it is to study the expectation values of Wilson loops in other geometries, e.g. squashed $S^3$, or lens spaces, or $S^2\times S^1$. A genuine quantum relation will hold for all geometries.  We will consider the squashed sphere in section \ref{sec:squa}.  Finally, isomorphism of the resulting algebras across known dualities is a strong consistency check, and we will verify this in section \ref{sec:dua}.

\subsection{Derivation of the Algebra from Matrix Model}

We will now attempt to derive these quantum relations by studying the matrix model which computes expectation values of BPS Wilson loops on $S^3$.  Consider Chern-Simons theory with a simple gauge group $G$ and level $k \in \ZZ$, with action:

\[ S_{CS} = \frac{k}{4 \pi} \int <A \wedge ,dA + \frac{2i}{3} A, \wedge [A \wedge ,A]> \]
where $<.,.>$ is the Killing form, normalized so that the longest roots have length $2$.  As described above, one can add auxiliary fields to ensure $\mathcal{N}=2$ supersymmetry, and compute the partition function by localization.  One finds that it is given by an integral over $\sigma$ in the Cartan $\mathfrak{h}$ of $G$:

\begin{equation}
\label{cspart}
Z_{G,k} = \frac{1}{|\mathcal{W}|} \int_{\mathfrak{h}} d\sigma e^{-k \pi i <\sigma,\sigma>} \prod_{\alpha \in Ad(G)} 2 \sinh \pi \alpha(\sigma)
\end{equation}
where $\alpha$ runs over the roots of the Lie algebra of $G$, and we divide by the order of the Weyl group $\mathcal{W}$.

One can also compute the expectation value of Wilson loops lying along fibers of the Hopf fibration, $\pi: S^3 \rightarrow S^2$.  The expectation value of a Wilson loop in a representation $R$ is given by an insertion in the integral of:

\begin{equation}
\label{wloi}
\mbox{Tr}_R e^{2 \pi \sigma}
\end{equation}
If there are multiple Wilson loops present, one simply includes one such factor in the integral for each of them.  Note that these Wilson loops have linking number $1$ with each other.

This implies that the product of two Wilson loops enters the matrix model in the same way as a single Wilson loop in the tensor product representation.  This is also true in general theories with matter.  Thus, at least at the level of expectation values, the product operation in the Wilson loop algebra is identical with the tensor product of representations.  Put another way, this suggests there is a natural homomorphism from the ring of representations of $G$ (the ring of formal linear combinations of representations of $G$ with product given by the tensor product) to the Wilson loop algebra, which we denote $\mathcal{A}$, sending a representation to a Wilson loop in that representation, and extending by linearity.  However, we will see in a moment that there are non-trivial relations satisfied by Wilson loop operators - i.e., this map has a non-trivial kernel - and so in general we expect the Wilson loop algebra to be some quotient of the representation ring.  

Thus it will first be important to discuss the representation ring of $G$ in more detail.  An alternative description of this ring is as the ring of class functions on $G$, i.e., functions which are constant on equivalence classes.  These functions are determined by their values on a maximal torus, or equivalently, as functions $f(\sigma)$ on the Cartan subalgebra $\mathfrak{h}$ of $\mathfrak{g}$ which can be lifted to functions on $T$.  Such functions take the form:

\begin{equation}
\label{rdef}
f(\sigma) = \sum_{\rho \in \Lambda} c_\rho e^{2 \pi \rho(\sigma)}
\end{equation}
here $\Lambda \subset \mathfrak{h}^*$ denotes the lattice of weights of $G$, and the $c_\rho$ are in the coefficient ring, which we presently take as $\ZZ$.  To be a class function, this expression must also be Weyl-symmetric.  It can be shown that an arbitrary such function can be written as a linear combination of traces of $e^{2 \pi \sigma}$ in irreducible representations of $G$, and the converse is also true, which gives a natural isomorphism between the two descriptions of $R(G)$.

Without risk of confusion we may then denote the set of such functions $f$ by $R(G)$.  Then we denote by $\tilde{R}(G)$ the ring of functions of the form (\ref{rdef}), but which are not necessarily Weyl-symmetric.  These are functions on the maximal torus which do not correspond to class functions.  We note that such functions can also be inserted into the matrix model, although since the rest of the integrand is Weyl-symmetric it will only receive a contribution from the symmetric part of $f$.

Let us now derive the relations determining the Wilson loop algebra as a quotient of $R(G)$.  For reasons that will become clear in a moment, it will actually be more natural to determine relations in $\tilde{R}(G)$, which determine an ideal $\mathcal{I}$.  We can then identify the Wilson loop algebra with the quotient of $R(G)$ by the Weyl-symmetric subideal of $\mathcal{I}$.

With this in mind, consider an insertion of $f \in \tilde{R}(G)$ into the matrix model.  Without loss, we can take $f=e^{2\pi \rho(\sigma)}$ for a chosen $\rho \in \Lambda$.  Then we find:

\begin{equation}
\label{wev}
< e^{2 \pi \rho(\sigma)} > = \frac{1}{|\mathcal{W}|} \int_{\mathfrak{h}} e^{-k \pi i <\sigma,\sigma>} e^{2 \pi \rho(\sigma)} \prod_{\alpha} 2 \sinh \pi \alpha(\sigma)
\end{equation}

Now the key point is to notice that the integrand in (\ref{wev}) is analytic as a function of $\sigma$, and has no poles in the complex $\sigma$ plane.  Thus it is invariant under suitably mild shifts of the integration contour.  In particular, consider shifting $\sigma \rightarrow \sigma + i x$, where $x$ lies in the coroot lattice, $\Lambda^*$.  Since this is dual to the weight lattice, we see $\rho(x) \in \mathbb{Z}$, and so this leaves the insertion $e^{2 \pi \rho(\sigma)}$ invariant.  It also preserves the product over sinh's.  The only change comes from the gaussian factor, and we are left with:

\begin{equation}
\label{wevshift}
 \int d \sigma e^{-k \pi i <\sigma,\sigma>} e^{k \pi i <x,x>} e^{2 \pi k \tilde{x}(\sigma)}  e^{2 \pi \rho(\sigma)} \prod_{\alpha} 2 \sinh \pi \alpha(\sigma)
\end{equation}
where we have defined $\tilde{x} \in \Lambda$ by:

\[ \tilde{x}( \sigma) = <x,\sigma> \]
Subtracting these two expressions, we find:

\[ < e^{2 \pi \rho(\sigma)} ( 1 -e^{k \pi i <x,x>}  e^{2 \pi k \tilde{x}(\sigma)} ) > = 0 \]
Since this is true for arbitrary $\rho$, and by linearity, for $e^{2\pi\rho(\sigma)}$ replace by an arbitrary $f \in \tilde{R}(G)$, we conjecture that the expression in parentheses actually vanishes in the algebra.  Thus we should impose the following relations on $\tilde{R}(G)$:

\begin{equation}
\label{rel}
1 - e^{k \pi i <x,x>} e^{2 \pi k \tilde{x}(\sigma)} = 0
\end{equation}
for arbitrary $x \in \Lambda^*$.

Let us briefly comment on the quantization of the Chern-Simons level $k$.  Given two coroots $x,y$, we find the relations:

\begin{equation}
1 - e^{k \pi i <x,x>}  e^{2 \pi k \tilde{x}(\sigma)} = 0, \;\;\;\; 1 - e^{k \pi i <y,y>} e^{2 \pi k \tilde{y}(\sigma)} = 0,
\end{equation}
Since $x+y$ is also a coroot, it also gives rise to a relation:

\begin{equation}
1 -e^{k \pi i <x+y,x+y>} e^{2 \pi k (\tilde{x}+\tilde{y})(\sigma)} = 0
\end{equation}
Comparing these, we see that together they imply:

\begin{equation}
\label{csquant}
e^{2k \pi i <x,y>} = 1
\end{equation}
Since $<x,y> \in \mathbb{Z}$, this reproduces the quantization condition $k \in \mathbb{Z}$.  This condition is also necessary in order for $k \tilde{x}$ to be a weight, in general.  Note that when this holds, the phase $e^{k \pi i <x,x>}$ is simply a sign.

 Let us denote the ideal of $\tilde{R}(G)$ generated by these relations by $\mathcal{I}'$.  Note that it is fixed by the action of the Weyl group on $\tilde{R}(G)$.  We then define the pre-Wilson loop algebra $\mathcal{A}'$ by:

\[ \mathcal{A}' =  R(G)/\mathcal{I}'_{\mathcal{W}} \]
where the subscript denotes taking the Weyl-symmetric subring of $\mathcal{I}'$.

There is one more set of relations we must impose to obtain the Wilson loop algebra $\mathcal{A}$.  Note that the integrand (\ref{wev}) contains a factor:

\[ V := \prod_{\alpha>0} (e^{2 \pi \alpha(\sigma)} - 1) \]
Note that $V$ is an element of $\tilde{R}(G)$.  Running through the argument above more carefully, we see the integral will vanish if $f V \in \mathcal{I}'$, even if $f$ itself is not.  Thus we should further divide by the set of elements in $\mathcal{A}'$ which give zero when multiplying $V$.\footnote{One might ask why we consider only $V$, when it is actually $V^2$ which appears in the integrand.  One can show in the examples of this section that all the elements which annihilate $V^2$ also annihilate $V$, and we conjecture this also holds in the general case.} This set forms an ideal called the annihilator of $V$, denoted $\mbox{Ann}_{\mathcal{A}'}(V)$.  Alternatively, we can think of this condition as defining a larger ideal $\mathcal{I}$ in $\tilde{R}(G)$.  Thus we define the Wilson loop algebra as:

\[ \mathcal{A} = \mathcal{A}'/\mbox{Ann}_{\mathcal{A}'}(V) \cong R(G)/\mathcal{I}_{\mathcal{W}} \]

This extra quotient actually has a simple consequence when we consider traces in irreducible representations.  Recall that once we pick a choice of positive roots, these representations are labeled by dominant weights $\rho$, and the trace in a representation with highest weight $\rho$  is given by the Weyl-character formula:

\begin{equation}
\label{wchar}
\mbox{Tr}_\rho (e^{2 \pi \sigma}) = \frac{A_{\delta+\rho}(\sigma)}{A_\delta(\sigma)}
\end{equation}
where $\delta$ is half the sum of the positive roots, and $A_\omega(\sigma) \in \tilde{R}(G)$ is defined for $\omega \in \mathfrak{h}^*$ as:

\[ A_\omega(\sigma) = \sum_{w \in \mathcal{W}} (-1)^w e^{2 \pi w \cdot \omega(\sigma)} \]
One can check that the denominator of (\ref{wchar}) divides the numerator in $\tilde{R}(G)$, and the resulting element is Weyl-symmetric, and lies in $R(G)$.  

The important point is that the denominator, $A_\delta$, is equal to $V$ (up to an overall monomial factor which does not affect the argument).  Then the division by the annihilator of $V$ is equivalent to imposing the relations (\ref{rel}) in the numerator of this expression.  In other words, we can write:

\[ \sum_{w \in \mathcal{W}} (-1)^w (e^{2 \pi w \cdot (\rho+\delta)(\sigma)} - (-1)^k e^{2 \pi (\rho + \delta + k \tilde{x})(\sigma)} )=  A_\delta( Tr_\rho e^{2 \pi \sigma} - (-1)^k Tr_{\rho + k \tilde{x}} e^{2 \pi \sigma} ) \in \mathcal{I}' \]
which is equivalent to:

\[ \Rightarrow Tr_\rho e^{2 \pi \sigma} - (-1)^k Tr_{\rho + k \tilde{x}} e^{2 \pi \sigma} \in \mathcal{I} \]
One gets such a relation for each dominant weight $\rho$.  To be precise, the weight $\rho + k \tilde{x}$ may not be dominant.  One can accommodate this by finding a Weyl transformation $w$ such that $\rho' + \delta:= w \cdot (\rho+\delta + k \tilde{x})$ is in the interior of the fundamental Weyl chamber, and replace the second term on the RHS by $(-1)^w Tr_{\rho'}$.  If no such $w$ can be found, i.e., if $\rho+\delta+k \tilde{x}$ lies on the boundary of a Weyl-chamber, then the expression vanishes by antisymmetry.

In fact, this last description of the algebra is precisely the standard description of the Verlinde algebra (see, e.g., \cite{2011arXiv1101.5887W}).  That is, one extends the ordinary Weyl group to the affine Weyl group by supplementing the reflections with translations by $k \rho$ for $\rho$ in the weight lattice, and one quotients out by these new relations as above.  The Verlinde algebra is known to be the algebra of Wilson loops in pure Chern-Simons theory, so it is encouraging that we reproduce this result.  

The arguments above may have been somewhat abstract, so let us illustrate them with a few examples.

\subsection{Examples}

\subsubsection{$U(N)$}

Consider the example $G=U(N)$.  This example is actually slightly outside the scope of the considerations above, since $U(N)$ is not simple, but the modifications are minor, and it will nevertheless illustrate the key ideas.  This will also be the main example we consider below when we add matter.

We parametrize the Cartan of $U(N)$ by $\sigma=\mbox{diag}(\lambda_1,...,\lambda_N)$.  The roots are given by $\alpha_{i,j}(\sigma) = \lambda_i - \lambda_j$ for $i \neq j$, and the Weyl group is the symmetric group $S_N$.  The Killing form is not unique, since the group is not semi-simple, but we can take it such that $<\sigma,\sigma> = \sum_j {\lambda_j}^2$.  In addition, because $U(N)$ has an $U(1)$ factor, one can include a Fayet-Iliopolous term $\zeta$.\footnote{This has a very simple effect in the pure Chern-Simons term: when integrating out the auxiliary fields, $\sigma$ is set to $\zeta$ rather than $0$, and the BPS Wilson loops are correspondingly weighted by powers of $e^{2 \pi\zeta}$ corresponding to their $U(1)$ charge.}  Then the partition function is given by:

\[ Z_{U(N)_k} = \frac{1}{N!} \int d^N \lambda e^{-k \pi i \sum_j {\lambda_j}^2 + 2 \pi i \zeta \sum_j \lambda_j} \prod_{i \neq j} 2 \sinh \pi (\lambda_i - \lambda_j) \]

Next we must discuss the representation ring and Wilson loop insertions.  The weights of $U(N)$ can be labeled by sequences $\rho=(\rho_1,...,\rho_N)$, with $\rho_j \in \ZZ$.  If we define $x_j = e^{2 \pi \lambda_j}$, then we see that $\tilde{R}(U(N))$ is nothing but the ring of Laurent polynomials in the $N$ variables $x_j$, and $R(U(N))$ is the ring of symmetric Laurent polynomials.  The latter is spanned by irreducible representations corresponding to dominant weights $\rho$, with $\rho_1 \geq ...\geq \rho_N$. One can label these by Young diagrams with $\rho_j$ boxes in the $j$th row.\footnote{We emphasize the $\rho_j$ may be negative - typically we only use the Young diagram description when they are all nonnegative.}

Now consider the insertion of $e^{2 \pi \rho(\sigma)} = e^{2 \pi \sum_j \rho_j \lambda_j}$ in the matrix model:

\[ <e^{2 \pi \rho(\sigma)}> = \frac{1}{N!} \int d^N \lambda e^{-k \pi i \sum_j {\lambda_j}^2+2 \pi i \zeta \sum_j \lambda_j} e^{2 \pi \sum_j \rho_j \lambda_j} \prod_{i \neq j} 2 \sinh \pi (\lambda_i - \lambda_j) \]
We will now perform the contour shifting argument as above.  We take as our coroots $x$ the vectors $x_\ell = \mbox{diag}(0,...,1,...,0)$, with all zeros except a $1$ in the $\ell$th position.  Shifting $\sigma \rightarrow \sigma +i x_\ell$ amounts to taking $\lambda_\ell \rightarrow \lambda_\ell+i$, and the integral above becomes:

\[ \frac{1}{N!} \int d^N \lambda e^{-k \pi i \sum_j {\lambda_j}^2+2 \pi i \zeta \sum_j \lambda_j} (-1)^k e^{-2 \pi \zeta } e^{2 \pi k \lambda_\ell} e^{2 \pi \sum_j \rho_j \lambda_j} \prod_{i \neq j} 2 \sinh \pi (\lambda_i - \lambda_j) \]
Subtracting these expressions, and noting this holds for general $\rho$, we are led to impose the following relations on $\tilde{R}(U(N))$:

\begin{equation}
\label{unrel}
 1 - (-1)^k e^{-2 \pi \zeta} e^{2 \pi k \lambda_\ell} = 1- (-1)^k q^{-1} {x_\ell}^k = 0, \;\; \ell=1,...,N 
\end{equation}
where we have defined $q:=e^{2 \pi \zeta}$.

At this point we should comment on the coefficient ring we are taking.  For $\zeta=0$, it is natural to simply take the integers, as described in section \ref{bpsw}.  For non-zero $\zeta$ we take the coefficient ring to be the ring of Laurent polynomials in the variable $q$.

Let us supposed $k>0$ - the case $k<0$ is analogous.  Then, given an arbitrary term $e^{2 \pi \rho(\sigma)} = \prod_j {x_j}^{\rho_j}$, applying this relation iteratively we can rewrite it as:

\begin{equation}
\label{unreps}
 ((-1)^k q)^n\prod_j {x_j}^{\rho_j'} \mbox{ where } 0 \leq \rho_j' < k \mbox{ and } \rho_j = \rho_j '\; (\mbox{mod} \; k) 
\end{equation}
where $n$ depends on the number of times the relation must be applied.  In particular, we can see the quotient algebra is finite dimensional over the relevant coefficient ring.

So far we have determined the ideal $\mathcal{I}'$, but we still need to quotient by the annihilator of $V$, which in this case is simply the usual Vandermone determinant (up to an irrelevant monomial factor):

\[ V = \prod_{i<j} (x_i - x_j) \]
As discussed above, it suffices to impose the relations (\ref{unrel}) on the Weyl-antisymmetric functions in the numerator of the Weyl character formula, and then divide by $V$ to obtain the inequivalent Weyl-symmetric functions.  In this case, an antisymmetric polynomial can be expanded in terms of polynomials of the form:

\[ \sum_{\pi} (-1)^\pi \prod_j {x_{\pi(j)}}^{\rho_j + N-j} \]
and, as in (\ref{unreps}), we may restrict to $\rho_j$ such that:

\[ 0 \leq \rho_N  < \rho_{N-1} + 1 < ... < \rho_1 + N-1 \leq k-1 \]

\[ \Rightarrow 0 \leq \rho_N \leq ... \leq \rho_1 \leq k-N \]
Such $\rho$ correspond to Young diagrams which fit into a box of size $N \times (k-N)$, and there are $\binom{k}{N}$ such choices.  In general the tensor product of such representations will decompose into representations which lie outside this box, but one can always use the relations to rewrite it in terms of representations which are contained in the box.  In particular, we see the dimension of the algebra is $\binom{k}{N}$, and in particular is finite.

\subsubsection{$Sp(2N)$}

As another example, we can take the group $Sp(2N)$.  We parameterize the Cartan of $Sp(2N)$ by:

\[ \sigma=\mbox{diag}(\lambda_1,...,\lambda_N,-\lambda_1,...,-\lambda_N) \]
The roots are given by:

\[ \alpha_{i,j,\pm,\pm}(\sigma) =  \pm \lambda_i \pm \lambda_j \mbox{ for } i < j \]

\[ \alpha_{i,\pm}(\sigma) = \pm 2 \lambda_i \]
The Weyl group is the semidirect product of the symmetric group $S_N$ with ${\ZZ_2}^N$, where the latter acts by flipping the signs of the $\lambda_j$.  The Killing form has $<\sigma,\sigma> = 2\sum_j {\lambda_j}^2$.  Then the partition function is given by:

\[ Z_{Sp(2N)_k} = \frac{1}{2^N N!} \int d^N \lambda e^{-2 k \pi i \sum_j {\lambda_j}^2} \prod_{i < j} (2 \sinh \pi (\lambda_i - \lambda_j))^2 (2 \sinh \pi (\lambda_i + \lambda_j))^2 \prod_j (-(2 \sinh (2 \pi \lambda_j))^2) \]

As above, the weights of $Sp(2N)$ can be labeled by sequences $\rho=(\rho_1,...,\rho_N)$, with $\rho_j \in \ZZ$.  $\tilde{R}(Sp(2N))$ is the ring of Laurent polynomials in the $x_j=e^{2 \pi \lambda_j}$, as above, but the Weyl group is different here, and $R(Sp(2N))$ is given by the Laurent polynomials which are symmetric both under exchanging the $x_i$ and taking $x_i \rightarrow {x_i}^{-1}$.  The dominant weights are those with $\rho_1 \geq ... \geq \rho_N \geq 0$, and we can again label these by Young diagrams.

Repeating the contour shifting argument as above, with $\lambda_\ell \rightarrow \lambda_\ell + i$, we find the relation:

\begin{equation}
\label{sprel}
1- e^{4 \pi k \lambda_\ell} = 1 - {x_\ell}^{2k}  = 0 
\end{equation}
As before, to determine a basis of the algebra, we consider Weyl-antisymmetric functions, which can be expanded in elements of the form:

\[ \prod_j {x_j}^{\rho_j + N+1 -j } + ... \]
where $\rho_1 \geq ... \geq \rho_N \geq  0$, and the dots represent terms arising from antisymmetrization.  Imposing the relation (\ref{sprel}) here, we see we can take all exponents to lie between $-k$ and $k$, so that $k-N-1 \geq \rho_1 \geq ... \geq \rho_N \geq 0$.  Thus the algebra is spanned by irreducible representations corresponding to Young diagrams which fit in a box of size $N \times (k-N-1)$, and has dimension $\binom{k-1}{N}$ over the coefficient ring $\ZZ$.

\section{Wilson Loop Algebra in Chern-Simons Theories with Matter}

We now extend the argument of the previous section to compute the algebra of BPS Wilson loops in Chern-Simons theories with matter.  By performing analogous manipulations of the matrix model as those that reproduced the Verlinde algebra above, we will find a simple generalization of this algebra that we argue is the correct algebra of BPS Wilson loops in these supersymmetric theories with matter.

We will start by illustrating the argument in a special case, that of $\cN=3$ $U(N)$ Chern-Simons theory with $N_f$ fundamental hypermultiplets, and then consider the general case of arbitrary gauge group and matter representation.

\subsection{$U(N_c)$ Theory with $N_f$ Fundamental Hypermultiplets}

Consider the theory with gauge group $U(N_c)$, a Chern-Simons term at level $k$, and $N_f$ fundamental hypermultiplets.  We assume the action preserves $\cN=3$ supersymmetry\footnote{This can be arranged by adding an adjoint chiral field, which does not contribute to the matrix model, along with a superpotential coupling this to the matter hypermultiplets.} and we also allow real masses $m_a$ for the $a$th hypermultiplet and a Fayet-Iliopolous term $\zeta$.  Then the partition function can be written as: 

\begin{equation}
\label{UNpart}
 Z= \frac{1}{N_c!} \int d^{N_c} \lambda \prod_{j=1}^{N_c} \frac{e^{- k \pi i {\lambda_j}^2 + 2 \pi \zeta \lambda_j}}{\prod_{a=1}^{N_f} 2 \cosh \pi (\lambda_j+ m_a)} \prod_{i \neq j } 2 \sinh \pi (\lambda_i - \lambda_j) 
\end{equation}
This differs from the expression (\ref{cspart}) for the Chern-Simons partition function by the inclusion of additional factors coming from the fluctuations of the matter fields around the BPS saddle points, labeled by $\sigma$.  In the $\cN=3$ case these enter as simple factors of $\cosh$, but in the more general case considered below they involve certain special functions.

A supersymmetric Wilson loop is again computed by an insertion of:

\[ \mbox{Tr}_R e^{2 \pi \sigma} \]

As before, we consider insertions from $\tilde{R}(U(N_c))$, and attempt to derive relations by manipulations in the matrix model.  Consider an insertion of $e^{2 \pi \rho(\sigma)}$ for a weight $\rho$:

\begin{equation}
\label{wevmat}
< e^{2 \pi \rho(\sigma)} > = \frac{1}{N_c!} \int d^{N_c} \lambda \prod_{j=1}^{N_c} \bigg( \frac{e^{-k \pi i {\lambda_j}^2 + 2 \pi i \zeta \lambda_j}}{\prod_{a=1}^{N_f} 2 \cosh \pi (\lambda_j + m_a)}  e^{2 \pi \rho_j \lambda_j} \bigg)  \prod_{i \neq j} 2 \sinh \pi (\lambda_i - \lambda_j)
\end{equation}

Now we would like to mimic the argument in the Chern-Simons case and use a deformation of the contour to derive relations between different Wilson loop insertions.  However, the factors of $\cosh$ in the denominator contribute poles at $\lambda_\ell = -m_a + (n+\frac{1}{2}) i $ for $n \in \ZZ$, and so the integral will no longer be invariant under a shift $\lambda_\ell \rightarrow \lambda_\ell + i $.  

There is a simple way around this problem, which is to insert additional factors into the integrand to cancel these poles.  In order to still derive relations in $\tilde{R}(U(N_c))$, these insertions must lie in $\tilde{R}(U(N_c))$ (with the appropriate coefficient ring).  Fortunately, there is a candidate insertion which does precisely this job, namely:

\[ \prod_{a=1}^{N_f} ( e^{2 \pi (\lambda_\ell + m_a)} + 1 ) \]
We see this cancels the appropriate factors of $\cosh$ in the denominator (up to a simple exponential factor), so that there are now no poles in the complex $\lambda_\ell$ plane.  Thus, with this insertion included, we are free to shift $\lambda_\ell \rightarrow \lambda_\ell+i$.  We obtain the relation:

\[ 
\frac{1}{N_c!} \int d^{N_c} \lambda \prod_{j=1}^{N_c} \frac{e^{-k \pi i {\lambda_j}^2 + 2 \pi i \zeta \lambda_j}}{\prod_{a=1}^{N_f} 2 \cosh \pi (\lambda_j + m_a)}  e^{2 \pi \sum_j \rho_j \lambda_j}   \prod_{a=1}^{N_f} ( e^{2 \pi (\lambda_\ell + m_a)} + 1 ) \prod_{i \neq j} 2 \sinh \pi (\lambda_i - \lambda_j) \]

\[ = \frac{1}{N_c!} \int d^{N_c} \lambda \prod_{j=1}^{N_c} \frac{e^{-k \pi i {\lambda_j}^2 + 2 \pi i \zeta \lambda_j}}{\prod_{a=1}^{N_f} 2 \cosh \pi (\lambda_j + m_a)}  (-1)^{k+N_f} e^{-2 \pi \zeta} e^{2 \pi k \lambda_\ell} e^{2 \pi \sum_j \rho_j \lambda_j}   \prod_{a=1}^{N_f} ( e^{2 \pi (\lambda_\ell + m_a)} + 1 ) \prod_{i \neq j} 2 \sinh \pi (\lambda_i - \lambda_j) \]
Subtracting these, and noting the result has zero expectation value for arbitrary $\rho$, we argue that the appropriate relations to impose here are:

\begin{equation}
\label{mattrel}
  ( 1 -(-1)^{k+N_f}  q^{-1} {x_\ell}^k) \prod_{a=1}^{N_f} ( x_\ell r_a + 1 ) =  0 , \;\;\;  \ell=1,...,N
\end{equation}
where we have defined $x_\ell=e^{2 \pi \lambda_\ell}$ and $q=e^{2 \pi \zeta }$, as above, as well as $r_a = e^{2 \pi m_a}$.  In this case we take the coefficient ring to be the polynomials in $r_a$ and $q$ which are symmetric in the $r_a$.

Note that this relation involves a polynomial in $x_\ell$ of degree $k+N_f$.  It can be rewritten in the form:

\[ {x_\ell}^{k+N_f} = \sum_{m=0}^{k+N_f-1} a_m {x_\ell}^m \]
for certain $a_m$ in the coefficient ring.  By iteratively applying this relation to a general monomial of the form $\prod_j {x_j}^{n_j}$, we can ensure that all the exponents $n_j$ lie in the range $0 \leq n_j \leq k+N_f-1$.

As before, these relations determine an ideal $\mathcal{I}'$.  Since the integrand still contains a factor of $V$, the same argument as in the previous section says that we must also quotient by the annihilator of $V$ to obtain the full idea $\mathcal{I}$ of vanishing Wilson loop operators.  As above, it suffices to consider antisymmetric functions, which can be expanded in terms of functions of the form:

\[ \sum_{\pi \in S_N} (-1)^\pi \prod_j {x_{\pi(j)}}^{\rho_j+N-j} + ... \]
and, using the relations (\ref{mattrel}), we can take $k+N_f-N_c \geq \rho_1 \geq ... \geq \rho_N \geq 0$.  Dividing by $V$ to obtain the characters of $U(N_c)$, we see that the algebra has a basis of irreps corresponding to Young diagrams which fit it a box of size $N_c \times (k+N_f-N_c)$, and the algebra has dimension $\binom{k+N_f}{N_c}$.   

In section \ref{sec:dua} we will present a very explicit description of this algebra in terms of generators satisfying specific relations.  

\subsection{Argument for General $\cN \geq 3$ Theories}

The argument above can be made quite general, and we can write down relations appropriate to a gauge theory with arbitrary gauge group and matter representations.  The case of theories with $\cN \geq 3$ supersymmetry is somewhat simpler, so we start here.

We consider a theory with a Chern-Simons kinetic term for the gauge group $G$, which we take to have the general form:

\[ S_{CS} = \frac{i}{4 \pi} \int \mbox{Tr}_{CS}( A \wedge dA + ... ) \]
for some trace functional $Tr_{CS}$ on the Lie algebra of $G$.  As in the case of pure Chern-Simons theory, it is necessary to impose a quantization condition $Tr_{CS}(xy) \in \mathbb{Z}$ for $x,y \in \Lambda^*$.  In addition, there is matter which comes in $M$ hypermultiplets, with the $a$th hypermultiplet living in a weight $\rho_a$ of the gauge group $G$ and $\tau_a$ of the flavor group $H$. Then we can parametrize the real masses by a parameter $\mu$ in the Cartan of $H$, and the partition function is given by:

\[ Z(\mu) = \frac{1}{|\mathcal{W}|} \int d \sigma e^{-i \pi Tr_{CS}(\sigma^2)} \prod_{a=1}^{M} \frac{1}{2 \cosh \pi (\rho_a(\sigma) + \tau_a(\mu))} \prod_{\alpha} 2 \sinh \pi\alpha(\sigma) \]
Here the fact that the matter comes in hypermultiplets and has $R$-charge $\frac{1}{2}$, as dictated by $\mathcal{N} \geq 3$ supersymmetry, guarantees the $1$-loop determinant is simply a product of $\cosh$'s.\footnote{In addition one could allow FI terms for $U(1)$ factors in the gauge group, as in the $U(N)$ case.  We do not consider such terms here, but it is a simple extension of the argument.}

As above, we consider an insertion of $e^{2 \pi \kappa(\sigma)} \in \tilde{R}(G)$ for a weight $\kappa$:

\[ <e^{2 \pi \kappa(\sigma)} >  =\frac{1}{|\mathcal{W}|} \int d \sigma e^{-i \pi Tr_{CS}(\sigma^2)} \prod_{a=1}^{M} \frac{1}{2 \cosh \pi (\rho_a(\sigma) + \tau_a(\mu))} e^{2 \pi \kappa(\sigma)} \prod_{\alpha} 2 \sinh \pi\alpha(\sigma) \]

Now we would like to shift $\sigma \rightarrow \sigma +i  x$.  Again we must include an insertion in $\tilde{R}(G)$ to cancel the poles.  These poles arise from those matter weights $\rho$ such that $\rho(x)$ is non-zero.  If we denote the set of such weights by $A_x$, consider the following insertion:

\[ <e^{2 \pi \kappa(\sigma)} \prod_{a \in A_x} (e^{2 \pi \rho_a(\sigma)} e^{2 \pi \tau_a(\mu)} + 1)> = \]

\[ =  \frac{1}{|\mathcal{W}|} \int d \sigma e^{-i \pi Tr_{CS}(\sigma^2)} \prod_{a=1}^{M} \frac{1}{2 \cosh \pi (\rho_a(\sigma) + \tau_a(\mu))} e^{2 \pi \kappa(\sigma)} \prod_{a \in A_x} (e^{2 \pi \rho_a(\sigma)} e^{2 \pi \tau_a(\mu)} + 1)\prod_{\alpha} 2 \sinh \pi\alpha(\sigma) \]
Now we see there are no poles when we shift $\sigma \rightarrow \sigma+ix$, so this integral is equal to:

\[  \frac{1}{|\mathcal{W}|} \int d \sigma e^{-i \pi Tr_{CS}(\sigma^2)} \prod_{a=1}^{M} \frac{1}{2 \cosh \pi (\rho_a(\sigma) + \tau_a(\mu))} \prod_{\alpha} 2 \sinh \pi\alpha(\sigma) \times \]

\[ \times  (-1)^{|A_x|} e^{\pi i Tr_{CS}(x^2)} e^{2 \pi Tr_{CS}(x \sigma)} e^{2 \pi \kappa(\sigma)} \prod_{a \in A_x} (e^{2 \pi \rho_a(\sigma)} e^{2 \pi \tau_a(\mu)} + 1) \]
Subtracting these and arguing as before, we found the following relation in $\tilde{R}(G)$:

\[  (1 - (-1)^{|A_x|} e^{\pi i Tr_{CS}(x^2)} e^{2 \pi Tr_{CS}(x \sigma)} )\prod_{a \in A_x} (e^{2 \pi \rho_a(\sigma)} e^{2 \pi \tau_a(\mu)} + 1) = 0\]

Because of the quantization condition we have imposed on $\mbox{Tr}_{CS}$, we see that $\mbox{Tr}_{CS}(x \sigma)$, as a linear functional of $\sigma$, is actually a weight.  In addition, it implies the factor $e^{\pi i Tr_{CS}(x^2)}$ is simply a sign.  The coefficient ring consists of sums of terms of the form $e^{2 \pi \tau(\mu)}$, and so is essentially the representation ring of the flavor group $H$.  As in the previous section, after imposing these relations we must also quotient by the annihilator of $V$.

We should emphasize that, for general theories, although we expect the relations above to hold, it is not clear that they are the only relations that need to be applied.  In the cases of $U(N)$ and $Sp(2N)$ theories with fundamental matter, which we consider in detail below, consistency with known dualities gives additional evidence that applying these relations alone gives the correct answer, as we will see in section \ref{sec:dua}.

\subsection{Algebra for $\cN=2$ Theories}
\label{n2alg}

The case of general $\cN=2$ theories is more complicated because the contribution of a chiral multiplet is given by a certain special function, the hyperbolic gamma function $\Gamma_h(z;\omega_1,\omega_2)$.  For $\omega_1/\omega_2 \in \mathbb{R}_{\geq 0}$, which will be the case of interest to us here, we can write \cite{VanDeBult2008}:

\[ \Gamma_h(z;\omega_1,\omega_2) = \exp\bigg( \pi i \frac{(2 z - \omega_1 -\omega_2)^2}{8 \omega_1 \omega_2} - \pi i \frac{{\omega_1}^2 + {\omega_2}^2}{24 \omega_1 \omega_2} \bigg) \prod_{j=0}^{\infty} \frac{1-e^{-2 \pi i \frac{z-(j+1)\omega_2}{\omega_1}}}{1-e^{-2 \pi i \frac{z+j\omega_1}{\omega_2}}} \]

Then the $S^3$ partition function of a chiral multiplet of $R$-charge $R$ and real mass $m$ can be written:

\[ \Gamma_h(i R + m;i,i) \]
The second and third arguments of the hyperbolic gamma function, which we will suppress for the rest of this section, are set to $i$ on the round sphere.  They attain more general values on the squashed sphere, which we will consider in section \ref{sec:squa}.

Now take a general theory with gauge group $G$ and flavor symmetry group $H$.  We assume there are $M$ chiral multiplets, and that the $a$th chiral lives in the weight $\rho_a$ of $G$ and $\tau_a$ of $H$ and has $R$-charge $R_a$.  We will assume the $R$-charges of all charged matter  lie between $0$ and $1$.  Then if we define real mass parameters $\sigma$ and $\mu$ which parameterize the Cartan of $G$ and $H$, respectively, its contribution is:

\[ \Gamma_h(i R_a + \rho_a(\sigma) + \tau_a(\mu) ) \]

Each chiral multiplet corresponds to a single pair $(\rho_a,\tau_a)$, and a product is taken over all of them.  This is inserted into the matrix model from the previous section, where we also allow a Chern-Simons term with trace $\mbox{Tr}_{CS}$.  We will comment on the quantization condition for this trace below.  The partition function is then given by:

\[ Z = \frac{1}{|\mathcal{W}|} \int d \sigma e^{-\pi i Tr_{CS} (\sigma^2)}  \prod_{a=1}^M \Gamma_h(i R_a + \rho_a(\sigma) + \tau_a(\mu)) \prod_{\alpha \in Ad(G)} 2 \sinh \pi \alpha(\sigma) \]

As always, Wilson loops correspond to insertions of characters of representations.  If we take an insertion $e^{2 \pi \kappa(\sigma)} \in \tilde{R}(G)$, we find:

\begin{equation}
\label{matloop}
< e^{2 \pi \kappa(\sigma)} > = \frac{1}{|\mathcal{W}|} \int d \sigma e^{- \pi i Tr_{CS} (\sigma^2)} e^{2 \pi \kappa(\sigma)} \prod_{a=1}^M \Gamma_h(i R_a + \rho_a(\sigma) + \tau_a(\mu)) \prod_{\alpha \in Ad(G)} 2 \sinh \pi \alpha(\sigma) 
\end{equation}
As in the previous two sections, we would like to derive relations by shifting $\sigma \rightarrow \sigma + i x$, for $x \in \Lambda^*$ after inserting an appropriate factor to cancel the poles in the integrand.  To do this, we will need to understand a few properties of the hyperbolic gamma function.

First, $\Gamma_h(z;i,i)$ has a pole of order $n+1$ at $z=-ni$ for all $n \in \ZZ_{\geq 0}$.  Thus the contribution of the chiral multiplet:

\[ \Gamma_h(i R + \rho(\sigma) + \tau(\mu)) \]
may cross poles when we shift $\sigma \rightarrow \sigma+ i x $.  Specifically, this can only occur when $\rho(x)<0$, since otherwise the imaginary part of the argument is always positive, noting $R>0$.  In fact, since $R<1$, we always will hit a pole whenever $\rho(x)<0$, so in these cases we must insert a factor to cancel these poles.\footnote{Actually, one should be careful that the poles are not canceled by zeros coming from other factors of $\Gamma_h(z)$.  One can show that this can only happen when the matter comes in hypermultiplet representations of $R$-charge a positive integer, so this will not concern us here.}

We will also need the fact that the hyperbolic gamma function satisfies the following difference equation (for general $\omega_i$):

\[ \Gamma_h(z+\omega_1;\omega_1,\omega_2) = 2 \sin \bigg( \frac{\pi z}{\omega_2} \bigg) \Gamma_h(z;\omega_1,\omega_2) \]
In the case $\omega_1=\omega_2=i$, this gives:

\[ \Gamma_h(z + i ) = -2 i \sinh (\pi z) \Gamma_h(z) \]
or, iterating this relation and rewriting it for later convenience, we get, for arbitrary $n \in \ZZ$:

\begin{equation}
\label{ghshift}
\Gamma_h(z + n i ) = e^{-\pi n z } e^{-\frac{\pi i n^2}{2}} ( e^{2 \pi z} - 1)^n \Gamma_h(z)
\end{equation}

With this in mind, let us fix some $x$ in the coroot lattice, and partition the set of chiral multiplets into subsets $A_+,A_-$, and $A_o$, corresponding to those with $\rho_a(x)$ positive, negative, and zero respectively.  Then, instead of (\ref{matloop}), we will consider the following integral:

\begin{equation}
\label{shifsta}
\int d \sigma e^{- \pi i Tr_{CS} (\sigma^2)} e^{2 \pi \kappa(\sigma)}  \bigg( \prod_a\Gamma_h(i R_a + \rho_a(\sigma) + \tau_a(\mu))\bigg) \bigg( \prod_{\alpha} 2 \sinh \pi \alpha(\sigma) \bigg)  \bigg( \prod_{a \in A_-} ( e^{-2 \pi (i R_a + \rho_a(\sigma) + \tau_a(\mu)} - 1)^{-\rho_a(x)} \bigg)
\end{equation}
Note the exponent in the final factor is positive, so this differs from (\ref{matloop}) by an insertion of an additional element of $\tilde{R}(G)$.  By (\ref{ghshift}), we see that it can be rewritten as:

\[ \int d \sigma e^{- \pi i Tr_{CS} (\sigma^2)} e^{2 \pi\kappa(\sigma)} \prod_{\alpha } 2 \sinh \pi \alpha(\sigma)  \times \]

\begin{equation}
\label{intshift}
\times \prod_{a \in A_+ \cup A_o} \Gamma_h(i R_a + \rho_a(\sigma) + \tau_a(\mu)) \prod_{a \in A_-}  e^{\pi \rho_a(x) (i R_a + \rho_a(\sigma) + \tau_a(\mu))} e^{-\frac{\pi i {\rho_a(x)}^2}{2}} \Gamma_h(i R_a + \rho_a(\sigma - i x) + \tau(\mu) )
\end{equation}
In this form it is clear that we do not cross any poles in shifting $\sigma \rightarrow \sigma+ix$, since the arguments of the hyperbolic gamma functions always have positive imaginary part.  Performing this shift, we find:

\[ \int d \sigma e^{- \pi i Tr_{CS} (\sigma^2)} e^{ \pi i Tr_{CS}(x^2)} e^{2  \pi Tr_{CS}(x \sigma) } e^{2 \pi \kappa(\sigma)} \prod_{\alpha } 2 \sinh \pi \alpha(\sigma)  \times \]

\begin{equation}
\label{intshift2}
\times \prod_{a \in A_+ \cup A_o} \Gamma_h(i R_a + \rho_a(\sigma+ i x) + \tau_a(\mu)) \prod_{a \in A_-}  e^{\pi \rho_a(x) (i R_a + \rho_a(\sigma) + i \rho_a(x) + \tau_a(\mu))} e^{-\frac{\pi i {\rho_a(x)}^2}{2}} \Gamma_h(i R_a + \rho_a(\sigma) + \tau(\mu) )
\end{equation}

\[ = \int d \sigma e^{- \pi i Tr_{CS} (\sigma^2)} e^{ \pi i Tr_{CS}(x^2)} e^{2  \pi Tr_{CS}(x \sigma) } e^{2 \pi \kappa(\sigma)} \bigg( \prod_a \Gamma_h(i R_a + \rho_a( \sigma) + \tau_a(\mu)) \bigg)  \prod_{\alpha } 2 \sinh \pi \alpha(\sigma)  \times \]

\begin{equation}
\label{shiftfin}
\times \prod_{a \in A_+} e^{-\pi \rho_a(x)(i R_a + \rho_a(\sigma) + \tau_a(\mu))} e^{-\frac{\pi i {\rho_a(x)}^2}{2}} ( e^{2 \pi (i R_a + \rho_a(\sigma) + \tau_a(\mu))} - 1)^{\rho_a(x)} \prod_{a \in A_-} e^{\pi \rho_a(x) (i R_a + \rho_a(\sigma) + i \rho_a(x) + \tau_a(\mu))} e^{-\frac{\pi i {\rho_a(x)}^2}{2}}
\end{equation}
Subtracting (\ref{shiftfin}) from (\ref{shifsta}), we obtain the relation:

\[ \left< e^{2 \pi \kappa(\sigma)} \bigg(   \prod_{a \in A_-} ( e^{-2 \pi (i R_a + \rho_a(\sigma) + \tau_a(\mu)} - 1)^{-\rho_a(x)}  -   \zeta e^{2 \pi \kappa_x(\sigma)} \prod_{a \in A_+} ( e^{2 \pi (i R_a + \rho_a(\sigma) + \tau_a(\mu))} - 1)^{\rho_a(x)}  \bigg)\right> = 0\]
where we have defined:

\[ \kappa_x(\sigma) = Tr_{CS}(x \sigma) - \frac{1}{2} \sum_{a \in A_+} \rho_a(x) \rho_a(\sigma) +  \frac{1}{2} \sum_{a \in A_-} \rho_a(x) \rho_a(\sigma) \]

\[ \zeta = e^{ \pi i Tr_{CS}(x^2)} \prod_{a \in A_+} e^{-\pi \rho_a(x)(i R_a + \tau_a(\mu))} e^{-\frac{\pi i {\rho_a(x)}^2}{2}} \prod_{a \in A_-} e^{\pi \rho_a(x)(i R_a + \tau_a(\mu))} e^{-\frac{\pi i {\rho_a(x)}^2}{2}} \]
Since this holds for arbitrary $\kappa$, we conjecture the Wilson loops satisfy the following relation in $\tilde{R}(G)$:

\begin{equation}
\label{genn2rel}
\prod_{a \in A_-} ( e^{-2 \pi (i R_a + \rho_a(\sigma) + \tau_a(\mu)} - 1)^{-\rho_a(x)}  -   \zeta e^{2 \pi \kappa_x(\sigma)} \prod_{a \in A_+} ( e^{2 \pi (i R_a + \rho_a(\sigma) + \tau_a(\mu))} - 1)^{\rho_a(x)}  = 0
\end{equation}
For this to make sense, it is important that $\kappa_x$ be a weight.  This imposes a quantization condition, as above, but it may differ from the condition above if the matter is not in a self-conjugate representation.  This is the manifestation in the matrix model of the parity anomaly.  We will see this explicitly in an example below.

\subsection{Examples}

\subsubsection{$U(N_c)$ with $N_f$ fundamentals}

As an example, let us again consider the case of $G=U(N_c)$ with fundamental matter.  Suppose we have $N_1$ fundamental and $N_2$ antifundamental chiral multiplets, so that the flavor symmetry group is $SU(N_1) \times SU(N_2)$.  If we write elements of the Cartan of the gauge and flavor groups as:

\[ \sigma = \mbox{diag}(\lambda_1,...,\lambda_{N_c}), \;\;\; \mu_1 = \mbox{diag}(m_1,...,m_{N_1}), \;\;\; \mu_2 = \mbox{diag}(\tilde{m}_1,...,\tilde{m}_{N_2}) \]
then the weights of the chiral multiplets take the form:

\[ ( \rho_j(\sigma),\tau_a(\mu_1) , 0) = (\lambda_j,m_a,0), \;\;\; j=1,...,N_c,\;\; a=1,...,N_1 \]

\[ (- \rho_j(\sigma),0,\tilde{\tau}_a(\mu_2)) = (-\lambda_j,0,\tilde{m}_a), \;\;\; j=1,...,N_c, \;\;a=1,...,N_2 \]
For simplicity we assign all chiral multiplets the same $R$-charge $r$.  Then the partition function is given by:

\[ Z = \frac{1}{N_c!} \int d^{N_c} \lambda e^{-\pi i k \sum_j {\lambda_j}^2 + 2 \pi i \zeta \sum_j \lambda_j}  \prod_{j=1}^{N_c} \bigg( \prod_{a=1}^{N_1} \Gamma_h(i r + \lambda_j + m_a) \prod_{a=1}^{N_2} \Gamma_h(i r - \lambda_j + \tilde{m}_a) \bigg) \prod_{i \neq j} 2 \sinh \pi (\lambda_i - \lambda_j) \]

Now we compute the Wilson loop algebra.  If we take $x_\ell=\mbox{diag}(0,...,1,...,0)$, with a $1$ in the $\ell$th position, one finds:

\[ A_+ = \{ ( \rho_\ell ,\tau_a) , a=1,...,N_1 \} , \;\;\; A_- = \{ ( -\rho_\ell ,\tilde{\tau}_a) , a=1,...,N_1 \}  \]

\[ \kappa_{x_\ell}(\sigma) =  \bigg( k + \frac{N_2-N_1}{2} \bigg) \lambda_\ell \]
and, if we define $s_a = e^{-2 \pi (i r + m_a)}$ and $\tilde{s}_a = e^{2 \pi (i r + \tilde{m}_a)}$, the relation (\ref{genn2rel}) becomes (after rearranging slightly):

\[ \prod_{a=1}^{N_2} {\tilde{s}_a}^{-1/2} (\tilde{s}_a x_\ell - 1) -  e^{\pi i (k - \frac{N_1+N_2}{2})} q^{-1} {x_\ell}^{k + \frac{N_2 - N_1}{2}}  \prod_{a=1}^{N_1} {s_a}^{-1/2} (s_a x_\ell - 1)= 0 \]

Note that the quantity $k + \frac{N_2 - N_1}{2}$ must be an integer in order to interpret this as a relation in $\tilde{R}(U(N))$.  This also ensures the phase $e^{\pi i (k - \frac{N_1+N_2}{2})}$ is just a sign.  In particular, if $N_1+N_2$ is odd, the Chern-Simons level must be a half-integer.  This agrees with the quantization of the Chern-Simons level in a chiral theory.

If we specialize to the case $N_1=N_2=N_f$, and define $r_a=(s_a \tilde{s}_a)^{1/2}$ and $t_a=-(s_a /\tilde{s}_a)^{-1/2}$, corresponding to vector and axial mass parameters respectively, then the relation can be written:

\begin{equation}
\label{unn2rel}
\prod_{a=1}^{N_f} (r_a t_a x_\ell  + 1) -  (-1)^{k+N_f} q^{-1} {x_\ell}^{k}  \prod_{a=1}^{N_f} (r_a x_\ell + t_a) = 0 
\end{equation}
The $\mathcal{N}=3$ case corresponds to setting $t_a=1$.  We see this is again a polynomial relation of degree $k+N_f$.  Thus, by the same argument as before, we expect the algebra to have dimension $\binom{k+N_f}{N_c}$.  Note also that one can flow between these theories by giving one flavor a large positive axial mass, which reduces $N_f$ while increasing $k$ and preserving $N_c$.  This corresponds to taking one of the $t_a$ to zero, and one can see from (\ref{unn2rel}) that this has precisely the effect of decreasing $N_f$ while increasing $k$.\footnote{Most other limits of real mass parameters also have the effect of changing $N_c$, which is subtle to see at the level of the partition function, and also at the level of the polynomial relation.}

\subsubsection{$Sp(2N_c)$ with $2N_f$ fundamentals}

As another example, take $G=Sp(2N_c)$ with $2N_f$ fundamental chiral multiplets.  The flavor group is $SU(2N_f)$, and we take parameterize the Cartans by:

\[ \sigma = \mbox{diag}(\lambda_1,...,\lambda_N,-\lambda_1,...,-\lambda_N) , \;\;\; \mu = \mbox{diag}(m_1,...,m_{2N_f}) \]
The weights of the chirals are:

\[ (\rho_{j,\pm}(\sigma),\tau_a(\mu)) = \pm \lambda_j + m_a , \;\;\; j=1,...,N_c,\;\; a=1,...,2N_f \]
The partition function is given by:

\[ Z_{Sp(2N_c)_{k,2N_f}}(m_a)  = \frac{1}{2^{N_c} {N_c}!} \int d^{N_c} \lambda e^{-2 k \pi i \sum_j {\lambda_j}^2} \prod_{j=1}^{N_c} \prod_{a=1}^{2N_f} \Gamma_h(\pm \lambda_j + m_a) \times \]

\[ \times \prod_{i < j} (2 \sinh \pi (\lambda_i - \lambda_j))^2 (2 \sinh \pi (\lambda_i + \lambda_j))^2 \prod_j (-(2 \sinh (2 \pi \lambda_j))^2) \]

As above we take $x_\ell= \mbox{diag}(0,...,1,...,-1,...,0)$, and one finds:

\[ A_\pm = (\rho_{\ell,\pm},\tau_a), \;\;\; a=1,...,2N_f \]

\[ \kappa_{x_\ell} = 2k \lambda_\ell \]
And the relation (\ref{genn2rel}) gives:

\begin{equation}
\label{sprel}
\prod_{a=1}^{2 N_f} ( {x_\ell} - s_a ) - e^{2 \pi i (k-N_f)} {x_\ell}^{2k} \prod_{a=1}^{2 N_f} (s_a {x_\ell} -1) = 0\end{equation}
where we define $s_a = e^{2 \pi (i R_a + m_a)}$.  Note that, in principle, one can take $k$ and $N_f$ both half-integral.  However, in the cases we consider the difference $k-N_f$ is an integer, so the factor $e^{2 \pi i(k-N_f)}$ drops out.  Since this relation is a polynomial of degree $2(k+N_f)$, by an analogous argument as in the pure Chern-Simons case we see we can restrict to insertions with powers of $x_j$ between $-(k+N_f)$ and $k+N_f$, and the algebra has dimension $\binom{k+N_f-1}{N_c}$.  We will discuss an explicit presentation of this algebra in the next section.

\section{Dualities}
\label{sec:dua}

In the previous sections we have presented a conjecture for the BPS Wilson loop algebra in $\cN=2$ theories with a Chern-Simons term.  This was derived by noting that certain linear combinations of Wilson loops acted as the zero operator, in the sense that their expectation value in the presence of arbitrary additional insertions was zero.  This argument by itself is not completely conclusive.  First, one could imagine these operators are not really zero, despite having zero expectation value.  Second, there could be other relations that one must impose besides the ones we found.

In the case of pure Chern-Simons theory, we have seen that the algebra we reproduce is precisely the same as the Verlinde algebra, which is known to be the correct algebra for Wilson loop operators.  In the case of Chern-Simons matter theories, there is no independent computation of this algebra to compare to.  However, one strong test that the algebra is correct in some cases would be if it is isomorphic for theories which are conjectured to be dual.  This also can be seen as a test of these proposed dualities.  

Thus in this section we consider some dualities of Chern-Simons matter theories, namely the dualities of Giveon and Kutasov \cite{Giveon:2008zn}, and argue that the algebra we have found above is indeed isomorphic for the dual pair.  In addition, we will show that this isomorphism, which is essentially unique, gives a prescription for mapping Wilson loop operators from one theory to its dual, and that the expectation value of these Wilson loops, as computed by the matrix model, agrees across the duality.

\subsection{Explicit Presentation of $R(U(N))$ Quotient Algebras}

First it will be necessary to derive a more convenient description of some of the algebras we have found above.  We will mainly focus on the case $G=U(N)$, but also consider $Sp(2N)$ below.  We have seen that the ring $\tilde{R}(U(N))$ is just the ring of Laurent polynomials in the $N$ variables $x_j = e^{2 \pi \lambda_j}$, and $R(U(N))$ is the subring of symmetric Laurent polynomials.  In all the quotient rings we have studied above, the relations have the form:\footnote{In more general theories, with matter representations other than the fundamental, the relations are typically polynomials involving multiple variables at once.  Here the analysis is more complicated, and we do not consider such algebras in detail in this paper.}

\[ p(x_\ell) = 0, \;\;\;\;\; \ell = 1,...,N \]
where $p$ is some degree $M$ polynomial.  In addition, we quotient out by elements which annihilate $V$, the Vandermonde determinant.  

Let us denote the quotient of $R(U(N))$ by these relations as $\mathcal{A}_{p}^{(N)}$, for an arbitrary $p$.  Then in this section we will present an explicit presentation of $\mathcal{A}_{p}^{(N)}$, and demonstrate an (essentially unique) isomorphism:

\[ h: \mathcal{A}_{p}^{(N)} \rightarrow \mathcal{A}_{p}^{(M-N)} \]
Below we will relate this isomorphism to certain field theory dualities, and demonstrate that the map $h$ gives the correct mapping of Wilson loop operators across the dualities.

We first consider the ring of ordinary symmetric polynomials (with non-negative powers), which we denote $\mathcal{S}(N)$, and will generalize to Laurent polynomials in a moment.   We can construct generators for this ring as follows.  Following \cite{salamon}, let us define:

\[ \Phi(t) = \prod_j (1 + t x_j) \]
which is a polynomial in $t$ of order $N$, where the coefficients $\phi_i=\phi_i(x_1,...,x_N)$ of $t^i$ are called the  elementary symmetric functions.  In terms of representations of $U(N)$, these symmetric functions can be written as Young diagrams as:

\[ \Phi(t) = 1 + {\tiny \yng(1)} \; t + {\tiny \yng(1,1)} \; t^2 + ... \]
Similarly, we define:

\[ \Psi(t) = \prod_j (1 - t x_j)^{-1} = 1 + {\tiny \yng(1)} \; t + {\tiny \yng(2)}\; t^2 + ... \]
which is an infinite series in $t$ with coefficients which we denote $\psi_i$.  Note the relation:

\begin{equation}
\label{phipsi}
\Phi(-t) \Psi(t) = 1
\end{equation}
which means the $\phi_i$ and $\psi_i$ are not independent, but one set of functions can be solved for in terms of the other by solving the above equation order by order in $t$.  The ring $\mathcal{S}(N)$ can then be defined as the ring generated by the $\phi_i$ and $\psi_j$ subject to the relations above.

Now suppose we want to impose the polynomial relations we found above, in addition to the condition related to $V$, on the ring of symmetric polynomials.  As a warm-up, consider the polynomial:

\[ p_k(x) = x^k\]
This arises in pure Chern-Simons theory when we take the limit $q \rightarrow \infty$.  To define this quotient ring, let us define a new generating function $\Psi_{k-N}(t)$ by truncating $\Psi(t)$ at order $k-N$:

\[ \Psi_{k-N}(t) = 1 + t \psi_1 + ... + t^{k-N} \psi_{k-N} \]

Then we claim that, in the quotient ring $\mathcal{A}_{p_k}^{(N)}$, one has the relation:

\[ \Phi(-t) \Psi_{k-N}(t) = 1 \]
This is demonstrated in the appendix, as a consequence of a more general formula we will describe in a moment.

We conjecture that there are no further relations in the algebra  $\mathcal{A}_{p}^{(N)}$. That is, it can be defined as the algebra generated by the $N$ elements $\phi_i$ and the $k-N$ elements $\psi_j$ subject to the relation above.  Note that this relation is symmetric under exchange of $N$ with $k-N$ and mapping $\phi_i \rightarrow (-1)^i \psi_i$.  In fact, this duality of the ring of symmetric functions in the nilpotent variables $x_i$ is well known; in particular, this ring is isomorphic to the cohomology ring of the Grassmannian $G(N,k)$, which obviously has this $N \leftrightarrow k-N$ duality.  We will see a physical explanation for this correspondence in section \ref{sec:connection}.  This is also related to level-rank duality of $U(N)$ Chern-Simons theory, as we will discuss shortly.

Next we turn to the case of general $p$.  Let us start by defining:

\[ p(x) = x^M + a_1 x^{M-1} + ... + a_M , \;\;\;\; \tilde{p}(t) = t^M p(t^{-1}) := 1 + a_1 t + a_2 t^2 + ... \]
where, without loss we have assumed $p(x)$ is monic.\footnote{This was not true for the algebras of the previous section, but it can be arranged by dividing these polynomial relations by a unit in the coefficient ring, which has no effect on the resulting algebra.}  Note that $\tilde{p}(t)$ need not have degree $M$, e.g., in the case $p_k(x)=x^k$ above, $\tilde{p}(t)=1$.  Then let us define $\Psi_p(t)$ as the truncation of the product $\tilde{p}(t) \Psi(t)$ after $M-N$ terms, ie:

\[ \Psi_p(t) = \lbrack \tilde{p}(t) \Psi(t) \rbrack_{M-N} = 1 + ({\tiny \yng(1)} + a_1 ) t + ({\tiny \yng(2)} + a_1 {\tiny \yng(1)} + a_2 ) t^2 + ... +(...) t^{M-N}\]
We denote the coefficient of $t^i$ in $\Psi_p(t)$ by ${\psi_p}_i$, $i=1,...,M-N$.  In the case $p(x)=x^k$ above, this reproduces $\Psi_{k-N}$.  Then it is shown in the appendix that, in the quotient ring $\mathcal{A}_{p}^{(N)}$, one has the relation:

\[ \Phi(-t) \Psi_p(t) = \tilde{p}(t) \]
Moreover, as above, we conjecture that $\mathcal{A}_{p}^{(N)}$ is precisely the ring generated by the $\phi_i$, $i=1,...N$, and the ${\psi_p}_j$, $j=1,...,M-N$, subject to this relation.

In this form, we see the isomorphism $h$ above is given by defining:

\[ h(\phi_i) = (-1)^i {\psi_p}_i \]
and extending so that $\phi$ is a homomorphism.  This gives the claimed isomorphism between $\mathcal{A}_{p}^{(N)}$  and $\mathcal{A}_{p}^{(M-N)}$.

Finally, we note that although we imposed these relations on the ring of symmetric polynomials, as opposed to symmetric Laurent polynomials, the final result is the same in both cases for generic $p(x)$.  Namely, provided $a_M$ is invertible in the coefficient ring, which will be true in all cases we consider,\footnote{Actually, this does not hold in the example of pure Chern-Simons theory with $q=e^{2\pi \zeta}\rightarrow \infty$ considered above, but only in the strict $\zeta \rightarrow \infty$ limit, which is somewhat degenerate.} we can write:

\[ x^{-1} = -{a_M}^{-1} ( x^{M-1} + a_1 x^{M-2} + ... + a_{M-1} ) \]
Since the ring of Laurent polynomials only differs by adjoining elements ${x_j}^{-1}$, we see this difference disappears one we pass to the quotient ring.

The isomorphism above is related to certain dualities of quantum field theories, which we describe now.  

\subsection{Level-Rank and Giveon-Kutasov Dualities}

First we consider a duality between two pure Chern-Simons theories, called level-rank duality.  In the non-supersymmetric notation, it asserts the equivalence of $U(N)$ Chern-Simons theory at level $K$ with $U(K)$ Chern-Simons theory at level $-N$.  In the supersymmetric notation, this amounts to:

\[ U(N)_k \leftrightarrow U(k-N)_{-k} \]
The Wilson loop algebras in the first theory is $\mathcal{A}_{p_k}^{(N)}$ where:

\[ p_k(x) = 1 - (-1)^k q^{-1} x^k \]
In the second theory, the relation is:

\[ p_{-k}(x) = 1 - (-1)^k q^{-1} x^{-k} \]
But this agrees with $p_k$ if we make the replacement $x \rightarrow x^{-1}$.  Thus the algebra here is $\mathcal{A}_{p_k}^{(k-N)}$.  From the previous section, these are isomorphic, and the isomorphism exchanging them sends:

\[ \phi_j \rightarrow (-1)^j \hat{\psi}_j \]
where the hat denotes charge conjugation, i.e., replacing $x \rightarrow x^{-1}$, or equivalently, $\sigma \rightarrow -\sigma$.  This equivalence of the algebras was also shown by a similar argument in \cite{Witten:1993xi}, where they were related to the quantum cohomology rings of the Grassmannian $G(N,k)$.

Recall that $\phi_j$ is represented by a Young diagram with one column and $j$ rows, while $\psi_j$ is represented by a Young diagram with one row and $j$ columns.  More generally, a consequence of this mapping is that the representation corresponding to a general Young diagram $Y$ is mapped to the transpose diagram $Y^T$, with a factor of $(-1)^{|Y|}$, where $|Y|$ is the number of boxes in the diagram.  This is well known to be the correct rule for mapping for Wilson loops in level-rank duality, and is consistent with the result above that the relevant Young diagrams fit into a box of size $N \times (k-N)$.  The equality of the expectation values of dual Wilson loops is shown, e.g., in \cite{Kapustin:2010mh}.

Next we turn to the duality of Giveon and Kutasov.  This asserts the equivalence of the following two theories:

\begin{itemize}
\item The theory with gauge group $U(N_c)$, Chern Simons level $k$, and $N_f$ fundamental hypermultiplets $(Q_a,\tilde{Q}^a)$ of $R$-charge $r$, with no superpotential.  We may assume $k>0$.
\item The theory with gauge group $U(k+N_f-N_c)$ at level $-k$ and $N_f$ fundamental hypermultiplets $(q_a,\tilde{q}^a)$ of $R$-charge $1-r$.  In addition, there are $N_f^2$ uncharged chiral multiplets ${M_a}^b$, which couple via a superpotential:

\[ W = \sum_{a,b} q^a {M_a}^b \tilde{q}_b \]
\end{itemize}

These theories have an $SU(N_f) \times SU(N_f)$ flavor symmetry rotating the two sets of chiral multiplets, as well as a $U(1)_J$ topological symmetry.  The duality dictates that the $SU(N_f)$ flavor symmetries of the two theories are identified up to a charge conjugation, and the $U(1)_J$ symmetries are identified precisely.  Thus if we deform by real vector and axial masses $m_a$, and $\mu_a$, and an FI term $\zeta$, these are mapped to the corresponding parameters $-m_a$,$-\mu_a$, and $\zeta$.  One can see the level rank duality above as a special case of this with $N_f=0$.

From (\ref{unn2rel}), the polynomial relation for the electric theory is given by:

\[ p(x_\ell) = \prod_{a=1}^{N_f} (r_a t_a x_\ell  + 1) -  (-1)^{k+N_f} q^{-1} {x_\ell}^{k}  \prod_{a=1}^{N_f} (r_a x_\ell + t_a) = 0  \]
where $r_a = e^{2 \pi m_a}$, $t_a=e^{2 \pi (i (r-1/2) + \mu_a)}$, and $q = e^{2 \pi \zeta}$.  Thus the algebra of the electric theory is $\mathcal{A}_p^{(N_c)}$.  

For the magnetic theory, we first note that the extra uncharged mesons have no effect on the algebra.  The identification of flavor symmetries implies that the polynomial here is:

\[ p'(x_\ell) = \prod_{a=1}^{N_f} ({r_a}^{-1} {t_a}^{-1} x_\ell  + 1) -  (-1)^{k+N_f} q^{-1} {x_\ell}^{-k}  \prod_{a=1}^{N_f} ({r_a}^{-1} x_\ell + {t_a}^{-1}) = 0  \]
and the algebra is $\mathcal{A}_{\tilde{p}}^{(k+N_f-N_c)}$.  

If we replace $x_\ell \rightarrow {x_\ell}^{-1}$ in $p'(x_\ell)$, and multiply by $\prod_a r_a t_a x_\ell$, we recover precisely the polynomial $p(x_\ell)$ for the electric theory.  Thus there is an isomorphism:

\[ h:\mathcal{A}_p^{(N_c)} \rightarrow \mathcal{A}_{p'}^{(k+N_f-N_c)} \]

\[ h(\phi_i) = (-1)^i \hat{\psi_p}_j \]

As a simple example, we may we set all the mass parameters $r_a$ and $t_a$ to $1$ and take $q \rightarrow \infty$.  Then the polynomials become simply:

\[ p(x) = x^k (x+1)^{N_f} , \;\;\; \tilde{p}(t) = (1+t)^{N_f} \]
And then the mapping $h$ sends:

\[ 1 \rightarrow 1 , \;\;\;\;  {\tiny \yng(1)} \rightarrow - ( {\tiny \yng(1)} +  \binom{N_f}{1} \cdot 1 )  \]

\[  {\tiny \yng(1,1)} \rightarrow {\tiny \yng(2)} + \binom{N_f}{1} \cdot {\tiny \yng(1)} + \binom{N_f}{2} \cdot 1 \]

\[  {\tiny \yng(1,1,1)} \rightarrow - \bigg( \; {\tiny \yng(3)} + \binom{N_f}{1} \cdot {\tiny \yng(2)} + \binom{N_f}{2} \cdot {\tiny \yng(1)} + \binom{N_f}{3} \cdot 1 \; \bigg)  \]

\[... \] 
We see that it generalizes the level-rank duality rule of transposing the Young diagrams by adding additional terms with fewer boxes.  More generally, we can rewrite $p(x)$ as a monic polynomial via:

\[ p(x) \rightarrow  {x_\ell}^{k} \prod_{a=1}^{N_f} ( x_\ell  + {r_a}^{-1} {t_a}^{-1}) -  (-1)^{k+N_f} q  \prod_{a=1}^{N_f} ({t_a}^{-1} x_\ell + {r_a}^{-1} ) \]

\[ \Rightarrow \tilde{p}(t) = \prod_{a=1}^{N_f} ( 1  + {r_a}^{-1} {t_a}^{-1} t) -  (-1)^{k+N_f} q t^k \prod_{a=1}^{N_f} ({t_a}^{-1} + {r_a}^{-1} t) \]
Then the mapping of Wilson loops has the same form as above, but with the binomial coefficient replaced by certain Laurent polynomials in $r_a$, $t_a$, and $q$. 

It is worth commenting that, for a commutative algebra over the complex numbers, one can always find a basis where the structure constants are diagonal, and the isomorphism class of the algebra is determined only by the dimension and the number of non-zero elements on the diagonal.   Thus isomorphism of the Wilson loop algebras as algebras over the complex numbers is not a very powerful check, and indeed there would be many such distinct isomorphisms in this case.  However, we argued above that the duality map must be an isomorphism of algebras over the ring $\CC[r_a,r_a^{-1},q, q^{-1}]$.  This is a much stronger constraint, and there is typically a unique such isomorphism.  In a moment we will demonstrate that the expectation values of dual Wilson loops in the ``magnetic theory'', mapped according to this prescription, are actually equal to expectation values in the ``electric'' theory. This provides an additional strong evidence that we have correctly identified the duality map.

\subsection{Consistency with Mapping of Defect Operators}

In \cite{Kapustin:2012iw} it was shown that Wilson loops in $U(N)$ theories which are abelian (i.e., in representations which only count the overall $U(1)$ charge) can be alternatively interpreted as defect operators in the $U(1)_J$ flavor symmetry of the theory.  To review the argument, we write the partition function with the $U(1)_J$ current for a vector multiplet $V$ coupled to a background vector multiplet $V_J$ as:

\begin{equation}
\label{defelec}
Z[V_J] = \int \mathcal{D} \Phi e^{i S[\Phi] + \frac{i}{2 \pi} \int d^3 x d^4 \theta V_J Tr \Sigma_V } 
\end{equation}
Then, given a choice of loop $\gamma$, there is a BPS vector multiplet ``vortex'' configuration, which we denote $\Omega_\gamma$, for which $A_\mu$ has a constant holonomy about $\gamma$, and $D$ is a delta function supported at $\gamma$.  Then one finds:

\[ Z[V_J + n \Omega_\gamma ] = \int \mathcal{D} \Phi e^{i S[\Phi] + \frac{i}{2 \pi} \int d^3 x d^4 \theta (V_J + n \Omega_\gamma ) Tr \Sigma_V }  = \int \mathcal{D} \Phi e^{i S[\Phi] + \frac{i}{2 \pi} \int d^3 x d^4 \theta V_J Tr \Sigma_V } e^{i n \int_\gamma Tr( A - i\sigma d|x|) }    \]
which is the same as a charge $n$ abelian Wilson loop insertion.

In the case of Giveon-Kutasov duality, the $U(1)_J$ symmetry of one theory maps to that of its dual.  Therefore, applying the construction above to both sides of the duality, one expects the charge $n$ abelian Wilson loop to map to the charge $n$ abelian Wilson loop on the dual side.  Actually, one must be slightly more careful because of the presence of contact terms.  Specifically, if we think of (\ref{defelec}) as giving the partition function of the ``electric'' theory, then, as argued in \cite{Closset:2012vp}, one must insert a level one Chern-Simons term for the gauge multiplet $V_{J}$ in the ``magnetic'' partition function:

\begin{equation}
\label{defmag}
\hat{Z}[V_J] = \int \mathcal{D} \Phi e^{i \hat{S}[\Phi] + \frac{i}{2 \pi} \int d^3 x d^4 \theta V_J Tr \Sigma_V + \frac{i}{4 \pi} \int d^3 x d^4 \theta V_J \Sigma_J} 
\end{equation}
This extra term is necessary for these partition functions to agree as functions of $V_J$.  Now when we perform the same shift $V_J \rightarrow V_J + n \Omega_\gamma$, we pick up additional terms:

\[ \hat{Z}[V_J + n \Omega_\gamma] = \int \mathcal{D} \Phi e^{i S[\Phi] + \frac{i}{2 \pi} \int d^3 x d^4 \theta V_J Tr \Sigma_V + \frac{i}{4 \pi} \int d^3 x d^4 \theta V_J \Sigma_J} e^{i n \int_\gamma Tr( A - i\sigma d|x|)} e^{i n \int_\gamma( A_J - i\sigma_J d|x|) } (-1)^n\]

Note that there is an additional charge $n$ Wilson loop for the background gauge field coupled to the $U(1)_J$ current, as well as a sign $(-1)^n$ which comes from a proper regularization of the phase proportional to ${\Omega_\gamma}^2$.\footnote{e.g., one can argue it is equivalent to smear the loop $\gamma$ uniformly over $S^3$, in which case this phase is easy to compute.}  At the level of the matrix model, this background Wilson loop enters simply as a factor of $e^{2 \pi n \zeta}=q^n$.

Now let us compare to the mapping of Wilson loops above.  For simplicity, let us restrict to the case $n=1$.  This corresponds to the representation we have labeled $\phi_N$ above, i.e., the Young diagram with one column with $N$ boxes.  Then the map of the previous section sends $\phi_N$ to ${\hat{\psi_p}_N}$ in the dual theory.  For this to agree with what we have just found, we must have the following relation in the dual:

\[ \hat{\psi_p}_N = -q \phi_{k+N_f-N} \]

However, note that the order $t^M$ term in the relation $\hat{\Phi}(-t) \hat{\Psi}_{p}(t) = \tilde{p}(t)$, which holds in this theory, gives:

\[ {\hat \phi}_{k+N_f-N} {\hat{\psi_p}}_{N} = -q  \]
In addition, one has $\phi_{k+N_f-N} {\hat \phi}_{k+N_f-N} = 1$ (i.e., already in the classical representation ring of $U(k+N_f-N)$), and so these relations are equivalent.

Note that the defect operator argument has suggested a new interpretation of the factor of $q$ as arising from a Wilson loop in the background gauge field coupled to the $U(1)_J$ symmetry.  By analogy, one might expect this also to be true of the factors of $e^{2 \pi m}$ that appear for real masses $m$ in generic flavor symmetries.

\subsection{Mapping of Wilson Loops in $\cN=3$ Giveon-Kutasov Duality}
\label{sec:wloopmap}
We now attempt to demonstrate that the isomorphism described above actually gives the correct mapping between Wilson loop operators across the Giveon-Kutasov duality by showing that the expectation values of the proposed dual operators are equal.  For the $\mathcal{N}=3$ version of the duality we will be able to present an analytic proof, while for the more general $\mathcal{N}=2$ case we can only check this numerically in several cases.

Recall the formula (\ref{UNpart}) for the partition function of the $\cN=3$ $U(N_c)$ theory with $N_f$ fundamental flavors.  It will be convenient to redefine this partition function by a phase so that it is real and positive.  As discussed in \cite{Closset:2012vp}, this phase can be attributed to certain contact terms that must be added to the action to ensure reflection positivity.  Thus we define:

\[ Z_{k,N_f,N_c}(\zeta,m_a) = e^{i \delta(N_c,k,N_f;\zeta,m_a)} \frac{1}{N_c!} \int d^{N_c} \lambda \prod_j \frac{e^{-k \pi i {\lambda_j}^2 + 2 \pi i \zeta \lambda_j}}{\prod_{a=1}^{N_f} 2 \cosh \pi (\lambda_j +m_a)} \prod_{i \neq j} 2 \sinh \pi (\lambda_i - \lambda_j) \]
where $\delta$ is chosen so that $Z$ is real and positive.  

Then the statement of Giveon-Kutasov duality is (defining $\hat{N}_c= k+N_f-N_c$):

\[ Z_{k,N_f,N_c}(\zeta,m_a) = Z_{-k,N_f,\hat{N}_c}(-\zeta,m_a) \]
There is also an explicit formula for the relative phase, which can be computed by studying the contact terms of the dual theories \cite{Closset:2012vp}:

\begin{align}
\label{phase}
\gamma(N_c,k,N_f;\zeta,m_a) := \delta(\hat{N}_c,-k,N_f;-\zeta,m_a)  - \delta(N_c,k,N_f;\zeta,m_a)  \notag \\ \notag \\
=  \frac{1}{24} (k^2 + 3 (k+N_f) (N_f-2) + 2) + \frac{1}{2} \zeta^2 - \frac{1}{2} k \sum_a {m_a}^2 - \zeta \sum_a m_a 
\end{align}

The isomorphism of the previous section gave an explicit prescription for how to map symmetric functions in the two algebras, namely, one exchanges $\Phi(-t)$ and $\Psi_p(t)$.  Thus the conjectured duality would imply that if one maps Wilson loops in the corresponding way, one should find that they have equal expectation values in the two theories, namely:

\begin{equation}
\label{abs} 
\left< \prod_\alpha \Phi(-t_\alpha) \prod_\beta \Psi_p(t_\beta) \right>_{N_c,k,N_f;\zeta,m_a} =  \left< \prod_\alpha \hat{\Psi}_p(t_\alpha) \prod_\beta \hat{\Phi}(-t_\beta) \right>_{\hat{N}_c,-k,N_f;-\zeta,m_a}
\end{equation}
where both sides are understood to include the appropriate phase factor (so that the partition function, given by $\left<1\right>$, is real and positive, although general Wilson loop expectation values will not be).  From this an arbitrary Wilson loop can be extracted by isolating the appropriate powers of $t_\alpha$ and $t_\beta$, since the $\phi_i$ and ${\psi_p}_j$ generate the algebra.

We will now prove (\ref{abs}) and (\ref{phase}).  This will provide a strong check both of the fact that $\mathcal{A}_{p}^{(N_c)}$ really is the Wilson loop algebra, and of the Giveon-Kutasov duality itself.  

The proof will be by induction on $N_f$.  We will not comment much on the case $N_f=0$, corresponding to level-rank duality, where the result is well known, and the mapping $\Phi(-t) \rightarrow \Psi_{k-N}(t)$ is understood as a special case of the flipping of the Young diagram rule for mapping Wilson loops.\footnote{See, e.g., \cite{Kapustin:2010mh} for a proof of the mapping of Wilson loop expectation values.}

The proof will be in two steps.  First we will express the general $N_f$-flavor Wilson loop expectation value in terms of a sum of $(N_f-1)$-flavor Wilson loop expectation values.  Then we will use induction to argue that the mapping of Wilson loops in the $(N_f-1)$-flavor theory gives rise to the desired mapping in the $N_f$-flavor theory.

Let us first introduce some notation.  We define:

\[ p(x) = (x^k - (-1)^{k+N_f} q) \prod_a (x + {r_a}^{-1}) \]

\[ p_b(x) = (x^k - (-1)^{k+N_f} q) \prod_{a \neq b} (x  + {r_a}^{-1}) \]
These are, respectively, the polynomials corresponding to the $(N_c,k,N_f)$ theory with masses $m_a$ and the theory with the $b$th flavor removed, for some arbitrary choice of $b$.\footnote{More precisely, we are using the wrong sign, $(-1)^{k+N_f}$, for the second theory; we will see why this is the appropriate choice in a moment.}

Then we claim an insertion of $\prod_\alpha \Phi(t_\alpha) \prod_\beta \Psi_p(t_\beta)$ in the $(N_c,k,N_f)$ theory can be expressed as:

\begin{align}
\label{formula}
 \left< \prod_\alpha \Phi(-t_\alpha) \prod_\beta\Psi_p(t_\beta) \right>_{N_c,k,N_f;\zeta,m_a} =&  \\
= \frac{1}{\tp_b(-r_b)} \bigg( e^{2 \pi i \alpha_1(N_c,k,N_f;\zeta,m_a;m_b)} & \left< \Psi_{p_b}(-r_b) \prod_\alpha \Phi(-t_\alpha) \prod_\beta(1 + t_\beta {r_b}^{-1}) \Psi_{p_b} (t_\beta) \right>_{N_c,k,N_f-1;\zeta-\frac{i}{2},m_a/m_b}  + \notag \\
 + e^{2 \pi i \alpha_2(N_c,k,N_f;\zeta,m_a;m_b)} &\left< \Phi(r_b) \prod_\alpha (1 + t_\alpha {r_b}^{-1}) \Phi(-t_\alpha) \prod_\beta \Psi_{p_b} (t_\beta)  \right>_{N_c-1,k,N_f-1;\zeta+\frac{i}{2},m_a/m_b} \bigg) \notag
\end{align}
This formula is proved in the appendix.  The basic idea of the proof is to perform a contour shift analogous to the one used to derive the algebra above, but without inserting extra factors to cancel the poles.  Then the term involving the $N_c-1$ theory appears from evaluating the residues at the poles that we cross.  Here $\alpha_i$ are some phase factors described in the appendix.  

Here, as always, we have assumed $k>0$.  We can modify this formula to work in general using:

\[ \left< \chi \right>_{N,-k,N_f;-\zeta,m_a} =  \bigg( \left< \hat{\chi} \right>_{N,k,N_f;\zeta^*,{m_a}^*} \bigg)^*\]
from which we can derive:

\[  \left< \prod_\alpha \hat{\Psi}_p(t_\alpha) \prod_\beta \hat{\Phi}(-t_\beta) \right>_{\hat{N}_c,-k,N_f;-\zeta,m_a} =\]

\[ =  \frac{1}{\tp_b(-r_b)} \bigg( e^{2 \pi i \alpha_1(\hat{N}_c,-k,N_f;-\zeta,m_a;m_b)}  \left< \hat{\Psi}_{p_b}(-r_b) \prod_\alpha (1+t_\alpha {r_b}^{-1})\hat{\Psi}_{p_b}(t_\alpha) \prod_\beta \hat{\Phi}(-t_\beta) \right>_{\hat{N}_c,-k,N_f-1;-\zeta-\frac{i}{2},m_a/m_b}  + \]

\[ + e^{2 \pi i \alpha_2(\hat{N}_c,-k,N_f;-\zeta,m_a;m_b)} \left< \hat{\Phi}(r_b) \prod_\alpha \hat{\Psi}_{p_b}(t_\alpha) \prod_\beta (1 + t_\beta {r_b}^{-1} ) \hat{\Phi} (-t_\beta)  \right>_{\hat{N}_c-1,-k,N_f-1;-\zeta+\frac{i}{2},m_a/m_b} \bigg)  \]
where we have extended the definition of $\alpha_i$ to negative $k$ by:

\[ \alpha_i(N_c,-k,N_f;-\zeta,m_a;m_b) = - \alpha_i(N_c,k,N_f;\zeta^*,m_a^*;m_b^*)^* \]

The formula (\ref{formula}) is derived in the appendix, along with the relation:

\begin{equation}
\label{phaserel}
\alpha_2(N_c,k,N_f;\zeta,m_a;m_b) = \alpha_1(\hat{N}_c,-k,N_f;-\zeta,m_a;m_b) 
\end{equation}
which automatically implies the corresponding relation with $1$ and $2$ exchanged.  Given these results, the proof of the mapping is straightforward.  Namely, by identifying terms in these two expressions, we see it would follow from:

\[ \left< \Psi_{p_b}(-r_b) \prod_\alpha \Phi(-t_\alpha) \prod_\beta(1 + t_\beta {r_b}^{-1}) \Psi_{p_b} (t_\beta) \right>_{N_c,k,N_f-1;\zeta-\frac{i}{2},m_a/m_b}  = \]

\[ = \left< \hat{\Phi}(r_b) \prod_\alpha \hat{\Psi}_{p_b}(t_\alpha) \prod_\beta (1 + t_\beta {r_b}^{-1} ) \hat{\Phi} (-t_\beta)  \right>_{\hat{N}_c-1,-k,N_f-1;-\zeta+\frac{i}{2},m_a/m_b}  \]
and:

\[ \left< \Phi(r_b) \prod_\alpha (1 + t_\alpha {r_b}^{-1}) \Phi(-t_\alpha) \prod_\beta \Psi_{p_b} (t_\beta)  \right>_{N_c-1,k,N_f-1;\zeta+\frac{i}{2},m_a/m_b} = \]

\[ = \left< \hat{\Psi}_{p_b}(-r_b) \prod_\alpha (1+t_\alpha {r_b}^{-1})\hat{\Psi}_{p_b}(t_\alpha) \prod_\beta \hat{\Phi}(-t_\beta) \right>_{\hat{N}_c,-k,N_f-1;-\zeta-\frac{i}{2},m_a/m_b}  \]
But these hold by induction on $N_f$.  Note that, although we are using the the sign $(-1)^{k+N_f}$ rather than $(-1)^{k+N_f-1}$ in the polynomial $p_b$ for these theories, this is corrected by the fact that $\zeta$ is shifted by $\pm \frac{i}{2}$, which changes $q$ by a sign.  This completes the proof of the mapping of Wilson loop expectation values across the duality.

\subsection{Giveon-Kutasov with an $Sp$ Gauge Group}

There is also a version of the Giveon-Kutasov duality with an $Sp$ gauge group, which reduces to the $Sp$ version of level-rank duality in the case $N_f=0$.  Here the dual theories are as follows:

\begin{itemize}
\item The theory with $Sp(2 N)$ gauge group, Chern Simons level $k$, and $2N_f$ fundamental chiral multiplets $Q_a$ of $R$-charge $R$, with no superpotential.
\item The theory with gauge group $Sp(k+N_f-N-1)$ at level $-k$ and $2N_f$ fundamental chiral multiplets $(q_a)$ of $R$-charge $1-R$.  In addition, there are $N_f^2$ uncharged chiral multiplets $M_{ab}$, which couple via a superpotential:

\[ W = \sum_{a,b} M^{ab} q_a q_b \]
\end{itemize}
As before, we consider an $\mathcal{N}=3$ version of the duality where the $R$-charges of the chiral fields are canonical and the $M$ fields are massive.

We now sketch how the arguments above are modified for the case $Sp(2N)$.  We first note that $\tilde{R}(Sp(2N)) =\tilde{R}(U(N))$, but the Weyl-symmetric condition is different in the two cases, as in the former case we must impose symmetry also under ${x_j} \rightarrow {x_j}^{-1}$, and so $R(Sp(2N)) $ is a proper subset of $R(U(N))$.  As before, we impose polynomial relations, which can be written as $p(x_j)=0$, where:

\[ x^{-k } \prod_{a=1}^{2 N_f} ( x^{1/2} {s_a}^{-1/2} - x^{-1/2} {s_a}^{1/2}) - x^{k} \prod_{a=1}^{2 N_f} ({s_a}^{1/2} x^{1/2} - {s_a}^{-1/2} x^{-1/2}) \]
where we have multiplied (\ref{sprel}) by an overall constant so that it has the form:

\[ q(x) - q(x^{-1}) \]
Note that $q$ is actually a polynomial, since $2N_f$ is even.  Let $M$ denote the degree of $q$.  

Now the argument is similar to the $U(N)$ case, although the details are somewhat different.  One defines, as above:

\[ \Phi(t) = 1 + {\tiny \yng(1)} \; t + {\tiny \yng(1,1)} \; t^2 + ... \]

\[ \Psi(t) = \prod_j (1 - t x_j)^{-1} (1 - t {x_j}^{-1})^{-1} = 1 + {\tiny \yng(1)} \; t + {\tiny \yng(2)}\; t^2 + ... \]
Here the Young diagrams represent representations of $Sp(2N)$, as computed by the Weyl character formula applied to the weight $(\rho_1,...,\rho_{N})$ with $\rho_j$ giving the number of boxes in the $j$th row of the diagram.  We note that $\Psi(t)$ still has a simple analytic form, but this is not true for $\Phi(t)$.

Now we consider imposing the relation $q(x_j) - q({x_j}^{-1}) = 0$, as well as the appropriate Vandermonde relation.  Then, if we define:

\[ \Psi_p(t) = \lbrack q(t) \Psi(t) \rbrack_{M-N-1} \]
where the brackets denote truncation after the $M-N-1$th term.  We conjecture the algebra is generated by the coefficients of $\Phi(t)$ and $\Psi_p(t)$ subject to the relation:

\[ (t^{N+1} \Phi(t^{-1}) - t^{-N-1} \Phi(t)) ( t^{M-N} \Psi_p(-t^{-1}) - t^{M-N} \Psi_p(-t)) = (-1)^M (q(-t) - q(-t^{-1}) ) ( t - t^{-1}) \]
This relation can be proven by a method similar to that in the $U(N)$ case, but again we do not know how to show that there are no further relations.  In particular, we see there is again an isomorphism sending:

\[ \phi_j \rightarrow (-1)^i {\psi_p}_j \]
This rule is very similar to the one for the $U(N)$ duality, with the Young diagrams mapping to transposed diagrams plus additional terms with fewer boxes.  One can check numerically for several low values of $N,k,N_f$ that this mapping preserves the expectation value of Wilson loops.

\section{Wilson loops on a squashed sphere}
\label{sec:squa}

To test our proposal for the algebra of Wilson loops further, we may replace the round $S^3$ with a squashed $S^3$, defined as a hypersurface in $\CC^2$ with the standard metric whose equation is
$$
\frac{1}{b^2}|z_1|^2 +b^2 |z_2|^2=1.
$$
One can parametrize the squashed sphere as follows:
$$
z_1=b\cos\theta e^{i\phi}, \quad z_2=b^{-1}\sin\theta e^{i\chi},
$$
where $\phi$ and $\chi$ have period $2\pi$ and $\theta$ takes values in the interval $[0,\pi/2]$. Thus a squashed $S^3$ can be thought of as a rectangular torus fibered over the interval $[0,\pi/2]$. At the ends of the interval the torus degenerates to a circle.

It was shown in \cite{Hama:2011ea} that the path-integral on a squashed $S^3$ for any $\cN=2$ $d=3$ theory localizes, and that matrix integral is written in terms of the double sine function, or equivalently, the hyperbolic gamma function.  For simplicity, we will consider in this section only the theory with $U(N)$ gauge group with $N_f$ fundamental hypermultiplets and Chern Simons level $k$ - the general case is a straightforward extension.  Then the partition function is given by:

\[ Z = \frac{1}{N!} \int d^N \lambda e^{\frac{\pi i k}{\omega_1 \omega_2} \sum_j {\lambda_j}^2 - \frac{2 \pi i \zeta}{\omega_1 \omega_2} \sum_j \lambda_j } \prod_{a=1}^{N_f} \prod_j  \Gamma_h( \omega R  \lambda_j + m_a +\mu_a;\omega_1,\omega_2) \Gamma_h( \omega R - \lambda_j - m_a +\mu_a);\omega_1,\omega_2) \times \]

\[ \times \prod_{i<j} 4 \sinh \pi b (\lambda_i - \lambda_j) \sinh \pi b^{-1} (\lambda_i - \lambda_j) \]
with $\omega=\frac{1}{2}(\omega_1+\omega_2)$.  On the squashed sphere, we have $\omega_1=i b,\omega_2=ib^{-1}$.

As in the case of round $S^3$, BPS Wilson loops must be wrapped along the integral lines of a vector field $v^\mu=\bar\eps\gamma^\mu\eps$ where $\eps$ is a solution of the twistor equation. For the choice of $\eps$ and $\beps$ used in \cite{Hama:2011ea}, this vector field is given by
$$
v^\mu\partial_\mu=b\frac{\partial}{\partial\chi}-b^{-1}\frac{\partial}{\partial\phi}.
$$
This vector field is tangent to the toroidal fibers, but its integral lines are not closed unless $b^2$ is rational.\footnote{For $\theta=0$ and $\theta=\pi/2$ the integral lines of $v^\mu$ are closed for all $b$. However, the vector field $v$ is not differentiable at  $\theta=0$ and $\theta=\pi/2$, therefore there might be subtleties with Wilson loops localized at these fibers. We will not consider them in this paper.} So let us assume that $b^2=m/n$ where $m$ and $n$ are relatively prime integers. Then for any $\theta$ and any initial point on the corresponding $T^2$ fiber the integral line is closed and wraps the $\chi$ and $\phi$ directions on the torus $m$ and $n$ times respectively. The length of this integral curve is independent of $\theta$ and is equal to $2\pi \sqrt {mn}=2\pi nb$. The corresponding BPS Wilson loop contributes a factor

\[ \mbox{Tr}_R(e^{2\pi i\frac{ n}{\omega_2}\sigma }) = \mbox{Tr}_R(e^{2\pi \sqrt{nm} \sigma }) \]
to the matrix model integrand.  Note that, topologically, these loops are $(n,m)$-torus knots.  In particular, we see that the matrix model allows one to compute the expectation value of torus knots on $S^3$ in Chern-Simons theory.

Using the scaling property $\Gamma_h( \alpha z;\alpha \omega_1,\alpha \omega_2)=\Gamma_h(z;\omega_1,\omega_2)$ with $\alpha=\sqrt{nm}$, and redefining $\sigma \rightarrow \frac{1}{\sqrt{nm}} \sigma$, and similarly for the other real mass parameters, we can rewrite the integrand above, with a Wilson loop insertion, as:

\[ <W_R> = \frac{1}{N!} \int d^N \lambda e^{-\frac{\pi i k}{n m} \sum_j {\lambda_j}^2 + \frac{2 \pi i}{nm} \zeta \sum_j \lambda_j }\mbox{Tr}_R(e^{2\pi \sigma })   \prod_{a=1}^{N_f} \prod_j  \Gamma_h( \frac{iR}{2}(m+n) \pm \lambda_j \pm m_a + \mu_a;i m, in) \times \]

\[ \times \prod_{i<j} 4 \sinh \frac{\pi}{m} (\lambda_i - \lambda_j) \sinh \frac{\pi}{n}(\lambda_i - \lambda_j) \]

We will now attempt to derive the quantum relations in the Wilson loop algebra.  We will proceed as above, starting with the case of pure Chern-Simons theory.  Here the integral, with an insertion $e^{2 \pi \kappa(\sigma)} = e^{2 \pi \sum_j \kappa_j \lambda_j}$ is:

\[ <e^{2 \pi \kappa(\sigma)} > = \frac{1}{N!} \int d^N \lambda e^{-\frac{\pi i k}{n m} \sum_j {\lambda_j}^2 + \frac{2 \pi i}{nm} \zeta \sum_j \lambda_j } e^{2 \pi \sum_j \kappa_j \lambda_j}\prod_{i<j} 4 \sinh \frac{\pi}{m} (\lambda_i - \lambda_j) \sinh \frac{\pi}{n}(\lambda_i - \lambda_j) \]
The analogous argument here would be to shift $\lambda_\ell \rightarrow \lambda_\ell+i n m$.  We see this leaves the gauge determinant and Wilson loop unchanged, up to a sign $(-1)^{(n+m)(N_c-1)}$, but the Chern-Simons and FI contributions change, and we find this is equal to:

\[ \frac{1}{N!} \int d^N \lambda e^{-\frac{\pi i k}{n m} \sum_j {\lambda_j}^2 + \frac{2 \pi i}{nm} \zeta \sum_j \lambda_j }(-1)^{k n m + (n+m)(N_c-1)} e^{-2 \pi \zeta } e^{2 \pi k \lambda_\ell} e^{2 \pi \sum_j \kappa_j \lambda_j}\prod_{i<j} 4 \sinh \frac{\pi}{m} (\lambda_i - \lambda_j) \sinh \frac{\pi}{n}(\lambda_i - \lambda_j) \]
Since this is true for arbitrary $\kappa$, we deduce the relation:

\[ 1 - (-1)^{k n m + (n+m)(N_c-1)} e^{-2 \pi \zeta } {x_\ell}^k= 0 \]
where we define $x_j = e^{2 \pi \lambda_j}$.  If we absorb the sign into a redefinition of the generators $x_j$, we see this is precisely the same relation as for the round sphere.  However, to argue for equivalence of the algebras, we must also impose the division by the annihilator of the Vandermonde determinant.  The argument used on the round sphere does not work in general here since the Vandermonde does not appear in the integrand (except for $n$ or $m$ equal to $1$).  

Nevertheless, although we were unable to find a general argument, we have checked in several cases by explicit computation (note the integral above is Gaussian) that the Vandermonde relations are indeed satisfied, and the algebra is precisely isomorphic to that on the round sphere.  This is not completely trivial, because the Wilson loops are topologically non-trivial, being $(n,m)$-torus knots.\footnote{Interestingly, the Vandermonde argument is straightforward when $n$ or $m$ is one, which are exactly the same cases where the knot is trivial.}  However, the Wilson loop algebra should depend only on local data, and not be sensitive to the topological nature of the knots.  Moreover, this ensures the algebra we find for these knots is consistent with level-rank duality.

Next let us consider adding matter.  For the round sphere, we started with the technically simpler $\mathcal{N}=3$ case where R-charges of all matter fields are $1/2$. For the squashed $S^3$ an analogous simplification occurs if we take

\[ r = \frac{\omega_2}{2 \omega}=\frac{n}{m+n}, \;\;\; \mu_a = 0 \]
in which case we find:

\[ \Gamma_h( \frac{i r}{2}(m+n) \pm \lambda_j \pm m_a + \mu_a;i m, in) = \frac{1}{2 \cosh \frac{\pi}{n} (\lambda_j + m_a) } \]
Thus the integral computing the Wilson loop expectation value can be written in terms of elementary functions:

\[ <W_R> = \frac{1}{N!} \int d^N \lambda \frac{e^{-\frac{\pi i k}{n m} \sum_j {\lambda_j}^2 + \frac{2 \pi i}{nm} \zeta \sum_j \lambda_j } }{  \prod_{a=1}^{N_f} 2 \cosh \frac{\pi}{n}(\lambda_j + m_a )} \mbox{Tr}_R(e^{2\pi \sigma })\prod_{i<j} 4 \sinh \frac{\pi}{m} (\lambda_i - \lambda_j) \sinh \frac{\pi}{n}(\lambda_i - \lambda_j) \]

We now shift $\lambda_\ell \rightarrow \lambda_\ell+i m n$ as above.  To avoid hitting poles from the matter we can insert a factor of:

\[ \prod_a ((-1)^n e^{2 \pi (\lambda_\ell + m_a)} - 1) \]
This factor cancels the poles in the denominator at $\lambda_\ell + m_a = n (s +\frac{1}{2})i$ for $s \in \mathbb{Z}$, and is also a valid Wilson loop insertion.  Accounting also for the extra contribution of the Chern-Simons term, we find the relation:

\begin{equation}
\label{n3rel}
 (1 - q (-1)^{k n m +m N_f  + (n+m) (N_c-1)} {x_\ell}^k) \prod_a ((-1)^n r_a x_\ell   -1) = 0 
\end{equation}
where we define $r_a = e^{2 \pi m_a}$.

Once again it is not clear that the Vandermonde relation holds in general here, but one can check this numerically in several cases and one finds it is always satisfied.  Then the algebra is a straightforward extension of the round sphere algebra, and by the general arguments of the previous section, is given by the algebra $\mathcal{A}_p(N_c)$ with:

\[ p(x) = (1 - q (-1)^{k n m +m N_f  + (n+m) (N_c-1)} x^k) \prod_a (x   - (-1)^n r_a )\]
The Giveon-Kutasov dual theory has algebra $\mathcal{A}_{p'}(k+N_f-N_c)$ with:\footnote{Note that the dual quarks have $R$ charge $1-r = \frac{m}{n+m}$, so their contribution to the matrix model is also a simple product of $\cosh$'s, although with $n \leftrightarrow m$ relative to the electric theory.}

\[ p'(y) = (1 - q (-1)^{k n m +n N_f  + (n+m) (k+N_f-N_c-1)} y^{-k}) \prod_a (y   - (-1)^m {r_a}^{-1} )\]
which, upon setting $y = (-1)^{n+m} x^{-1}$, agrees with the polynomial above, up to an overall monomial factor which does not affect the algebra, and so the algebras are isomorphic.  In fact, up to redefinitions of the parameters by signs, we see the algebra is independent of $n$ and $m$, as expected if the algebra is a property of the underlying theory, and not special to the round sphere. We have checked numerically in several cases that Wilson loop expectation values map as expected under this duality.

Finally we consider the general case, of arbitrary $R$ charge and axial mass.  Recall the Wilson loop in this case is computed by:

\[ <W_R> = \frac{1}{N!} \int d^N \lambda e^{-\frac{\pi i k}{n m} \sum_j {\lambda_j}^2 + \frac{2 \pi i \zeta}{n m} \sum_j \lambda_j } \mbox{Tr}_R (e^{2 \pi \sigma} )\prod_{a=1}^{N_f} \prod_j  \Gamma_h( \frac{n+m}{2} i R \pm( \lambda_j + m_a) +\mu_a;i n, i m) \times
\]

\[ \times \prod_{i<j} 4 \sinh \frac{\pi}{n} (\lambda_i - \lambda_j) \sinh \frac{\pi}{m} (\lambda_i - \lambda_j) \]

We will need to make use of the following difference equation satisfied by the functions $\Gamma_h(z)$:

\[ \Gamma_h(z + \omega_1) = 2\sin \bigg( \frac{ \pi z}{\omega_2} \bigg) \Gamma_h(z) \]
with a similar equation with $\omega_1 \leftrightarrow \omega_2$. 

Let us consider $\frac{\omega_1}{\omega_2}=\frac{m}{n}$ as before.  Then consider:

\[ \Gamma_h(z + n \omega_1) = \bigg(\prod_{j=0}^{n-1} 2\sin \bigg( \frac{ \pi (z + j \omega_1)}{\omega_2} \bigg)  \bigg) \Gamma_h(z) \]
Using the assumed form of $\frac{\omega_1}{\omega_2}$, the product on the left can be simplified, and this becomes:

\[ \Gamma_h(z + n \omega_1) = \zeta 2 \sin \bigg( \frac{\pi n z}{\omega_2} \bigg) \Gamma_h(z) \]
where $\zeta=e^{\frac{\pi i}{2} (m-1)(n-1)} $.  More generally, we find, for $p \in \ZZ$:

\[ \Gamma_h(z + p n \omega_1) = (-1)^{n m p (p-1)/2} \zeta^p \bigg( 2 \sin \bigg( \frac{\pi n z}{\omega_2} \bigg) \bigg)^p \Gamma_h(z)  \]
In the case $\omega_1=i m, \omega_2=i n$, this gives:

\begin{equation}
\label{phase1}
\Gamma_h(z + i p n m ) = (-1)^{n m p (p-1)/2} \zeta^p ( 2 \sinh ( \pi z ))^p \Gamma_h(z) 
\end{equation}
Now the argument proceeds much as on the round sphere.  We start with the expression:

\[ \frac{1}{N!} \int d^N \lambda e^{-\frac{\pi i k}{n m} \sum_j {\lambda_j}^2 + \frac{2 \pi i \zeta}{n m} \sum_j \lambda_j } \mbox{Tr}_R (e^{2 \pi \sigma} )\prod_{a=1}^{N_f} \bigg(\prod_{j \neq \ell}  \Gamma_h( \frac{n+m}{2} i R \pm( \lambda_j + m_a) +\mu_a;i n, i m) \bigg) \times
\]

\[ \times \gamma \prod_a \Gamma_h( \frac{n+m}{2} iR + ( \lambda_\ell + m_a) +\mu_a;i n, i m) \bigg) \Gamma_h( \frac{n+m}{2} iR - ( (\lambda_\ell- n m i) + m_a) +\mu_a;i n, i m) \bigg) \times \]

\[\times \prod_{i<j} 4 \sinh \frac{\pi}{n} (\lambda_i - \lambda_j) \sinh \frac{\pi}{m} (\lambda_i - \lambda_j) \]
where we have included a factor:

\[ \gamma = \prod_a e^{\pi( \frac{n+m}{2} i R -  (\lambda_\ell+ m_a) +\mu_a )} \]
for later convenience.  This manifestly crosses no poles upon shifting $\lambda_\ell \rightarrow \lambda_\ell + i n m$, and so is equal to:

\[ \frac{1}{N!} \int d^N \lambda e^{-\frac{\pi i k}{n m} \sum_j {\lambda_j}^2 + \frac{2 \pi i \zeta}{n m} \sum_j \lambda_j } \mbox{Tr}_R (e^{2 \pi \sigma} )\prod_{a=1}^{N_f} \bigg(\prod_{j \neq \ell}  \Gamma_h( \frac{n+m}{2} i R \pm( \lambda_j + m_a) +\mu_a;i n, i m) \bigg)  (-1)^{n m N_f+k n m + (n+m)(N_c-1)}   \times \]

\[ \times e^{-2 \pi i \zeta} e^{2 \pi k \lambda_\ell} \gamma \prod_a \Gamma_h( \frac{n+m}{2} i R + ( (\lambda_\ell+ n m i) + m_a) +\mu_a;i n, i m) \Gamma_h( \frac{n+m}{2} i R - ( \lambda_\ell+ m_a) +\mu_a;i n, i m) \bigg) \times \]

\[ \times \prod_{i<j} 4 \sinh \frac{\pi}{n} (\lambda_i - \lambda_j) \sinh \frac{\pi}{m} (\lambda_i - \lambda_j) \]
These expressions can be rewritten using (\ref{phase1}), and subtracting them we find:

\[0 = \frac{1}{N!} \int d^N \lambda e^{-\frac{\pi i k}{n m} \sum_j {\lambda_j}^2 + \frac{2 \pi i \zeta}{n m} \sum_j \lambda_j } \mbox{Tr}_R (e^{2 \pi \sigma} )\prod_{a=1}^{N_f} \bigg(\prod_j \Gamma_h( \frac{n+m}{2} i R \pm( \lambda_j + m_a) +\mu_a;i n, i m) \bigg)  \times
\]

\[   \gamma \bigg( \prod_a 2 \sinh \pi( \frac{n+m}{2} i R -(\lambda_\ell+ m_a) + \mu_a )- (-1)^{(k+N_f)n m + (n+m)(N_c-1)} e^{-2 \pi i \zeta} e^{2 \pi k \lambda_\ell} \prod_a  2 \sinh \pi( \frac{n+m}{2} i R +(\lambda_\ell+ m_a) + \mu_a ) \bigg)  \]

\[ \times \prod_{i<j} 4 \sinh \frac{\pi}{n} (\lambda_i - \lambda_j) \sinh \frac{\pi}{m} (\lambda_i - \lambda_j) \]
Using the form for $\gamma$ above, the second line can be rewritten as:

\[  \prod_a (e^{2 \pi( -\frac{n+m}{2} i R +\lambda_\ell+ m_a - \mu_a )} - 1) - (-1)^{(k+N_f)n m + (n+m)(N_c-1)} e^{-2 \pi i \zeta} e^{2 \pi k \lambda_\ell} \prod_a  ( e^{2 \pi( -\frac{n+m}{2} i R - \mu_a )} - e^{2 \pi (\lambda_\ell + m_a)} ) \]
which, upon defining $x_\ell=e^{2 \pi \lambda_\ell}$, $r_a=e^{2 \pi m_a}$, $s_a = e^{- 2 \pi(\frac{n+m}{2} i R + \mu_a)}$, and $q= e^{-2 \pi \zeta}$, implies the following relation in the algebra:

\[ p(x_\ell) :=  \prod_a ( s_a r_a x_\ell - 1) - (-1)^{(k+N_f)n m + (n+m)(N_c-1)} q {x_\ell}^k \prod_a  ( s_a - r_a x_\ell ) = 0 \]
Note that the special case above, $R=\frac{n}{n+m}$ and $\mu_a=0$, corresponds to setting $s_a=(-1)^n$, and we recover the relation (\ref{n3rel}).

Once again, we assume without proof that the Vandermonde condition is also satisfied in this case, in which case the algebra is simply $\mathcal{A}_{p}(N_c)$, with $p(x)$ as above.  For the dual theory, $U(k+N_f-N_c)_{-k}$ with $N_f$ flavors, we have $r_a \rightarrow {r_a}^{-1}$ and $s_a \rightarrow (-1)^{n+m} {s_a}^{-1}$, and so the polynomial is:

\[ p'(y) = \prod_a ( (-1)^{n+m} {s_a}^{-1} {r_a}^{-1} y - 1) - (-1)^{(k+N_f)n m + (n+m)(k+N_f-N_c-1)} q y^{-k} \prod_a  ( (-1)^{n+m} {s_a}^{-1} - {r_a}^{-1} y )  \]
Upon substituting $y=(-1)^{n+m} x^{-1}$, one finds this precisely agrees with the polynomial above (up to an overall factor), and so by the general arguments of the previous section, the algebras are dual.

Unfortunately, we do not know how to evaluate the integrals numerically for generic $R$ charge, so we were unable to perform numerical checks in this case.
\section{Connection with quantum cohomology and quantum K-theory}\label{sec:Ktheory}
\label{sec:connection}

In the case of pure Chern-Simons theory with $G=U(N_c)$, the algebra of Wilson loops can be computed in another way. When compactified on $S^1$, Chern-Simons theory becomes equivalent to the $G/G$ model (the gauged WZW model). On the other hand, E. Witten argued \cite{Witten:1993xi} that for $G=U(N_c)$ the $G/G$ model at level $\kb$ is equivalent to the A-model whose target is the complex Grassmannian of $N_c$-planes in $\CC^{|k|}$, where $k=\kb+N_c \sign\, \kb$.  Wilson loops wrapped on $S^1$ become local operators in the $G/G$ model, therefore the Wilson loop algebra for the bosonic $G=U(N_c)_\kb$ Chern-Simons theory must be isomorphic to the quantum cohomology ring of $Gr(N_c,\CC^{|k|})$. 

In this section we generalize this approach to $U(N_c)$ Chern-Simons theories with matter in the fundamental representation. We begin by deriving the relation between the Verlinde algebra and the quantum cohomology of Grassmannian using compactification of a suitable $\cN=2$ $d=3$ theory on a circle.\footnote{We are grateful to K. Hori for help with this argument, and especially for pointing out that the right starting point is a Chern-Simons theory at level $k/2$ and $k$ fundamentals.}

Consider $\cN=2$ $d=3$ gauge theory with gauge group $U(N_c)$ and $k$ chiral multiplets $q_A$, $A=1,\ldots,k,$ in the fundamental representation. The action for the vector multiplet is taken to be the $\cN=2$ Chern-Simons action at level $k/2$.  (For odd $k$ this level is half-integer, precisely as required by the cancellation of gauge anomalies). Let us analyze the vacuum structure of this model as a function of the FI parameter $\zeta$. First suppose $\zeta<0$. Let the scalar $\sigma$ have eigenvalues $\sigma_i$, $i=1,\ldots,N_c$. If $\sigma_i\neq 0$, vanishing of the potential requires $q^i_A=0$ for all $A$.. For such $i$ the D-flatness condition implies
$$
\zeta+\frac{k}{2}\sigma_i+\frac{k}{2} |\sigma_i|=0.
$$
The second term here is the contribution of the classical Chern-Simons term, while the third term is due to the shift of the Chern-Simons coupling arising from integrating $k$ charged chiral multiplets $q^i_A$, $A=1,\ldots,k$. For $\zeta<0$ this equation has a unique solution $\sigma_i=-\zeta/k>0$. Thus all nonzero eigenvalues of $\sigma$ are equal and positive. On the other hand, if for some $i$ we have $\sigma_i=0$, then the D-flatness condition implies
$$
\zeta=\sum_{A=1}^{k} q^{A \dagger}_i q^i_A.
$$
For $\zeta<0$ this equation has no solutions. We conclude that for $\zeta<0$ the theory has a unique vacuum where $\sigma=-\zeta/k$. In this vacuum the gauge group is unbroken and all the matter fields are massive and can be integrated out. This shifts the Chern-Simons level from $k/2$ to $k$. We conclude that this vacuum is described by an $\cN=2$ $U(N_c)$ Chern-Simons theory at level $k>0$ (and no matter fields). This theory exists provided $k\geq N_c$. For $k<N_c$ and $\zeta<0$ the theory must break supersymmetry spontaneously.

Now suppose $\zeta>0$. Then the same argument shows that all eigenvalues of $\sigma$ must vanish, and the D-flatness condition reads
$$
\zeta\delta^i_j=\sum_{A=1}^k q^i_A q^{A \dagger}_j.
$$
For $k\geq N_c$ the moduli space of solutions of this equation is the Grassmannian $Gr(N_c,\CC^k)$. Thus the low-energy theory is described by an $\cN=2$ sigma-model with target $Gr(N_c,\CC^{k})$. For $k<N_c$ there is no supersymmetric vacuum, and the theory breaks supersymmetry spontaneously. 

Now let us compactly this theory on a circle of radius $R$. This gives a 2d theory with $\cN=(2,2)$ supersymmetry. The effective 2d FI term depends on the RG scale $\mu$ and for $\mu$ much smaller than $1/R$  is (with logarithmic accuracy)
$$
\zeta_2=2\pi R\zeta+k \log (\mu R).
$$
Now, supersymmetry implies that the twisted chiral ring of this theory is determined by the IR physics alone.\footnote{Formally, this follows from the existence of a semi-topological twist of the $\cN=2$ $d=3$ theory compacted on a circle such that varying $R$ changes the action only by BRST-exact terms.} So if we fix $\mu$ and $\zeta_2$ and start varying $R$ and $\zeta$, we must find the same twisted chiral ring regardless of the value of $R$. Suppose $\zeta_2$ is large and negative. Then
$$
2\pi R\zeta=-k \log (\mu R e^{|\zeta_2|}).
$$
This expression can have either sign, depending on the value of $R$. If it is positive, the effective 2d theory at energy scale $1/R$ is the $\cN=(2,2)$ sigma-model with target $Gr(N_c,\CC^k)$. If it is negative, the effective 2d theory at energy scale $1/R$ is the compactification of the $\cN=2$ Chern-Simons theory at level $k$. Thus the algebra of Wilson loops in the latter theory is isomorphic to the twisted chiral ring of the Grassmannian sigma-model. Or equivalently, to the quantum cohomology algebra of $Gr(N_c,\CC^k)$.  This is the result which Witten obtained by analyzing the dynamics of the Grassmannian sigma-model \cite{Witten:1993xi}.

If we want to get the Verlinde algebra at negative level $k<0$, we just need to start with $\cN=2$ $d=3$ $U(N_c)$ gauge theory at level $k/2$ coupled to $|k|$ chiral multiplets in the anti-fundamental representation.  Repeating the above analysis, we find that for $\zeta>0$ the 3d theory flows to the $\cN=2$ $U(N_c)$ Chern-Simons theory at level $k$, while for $\zeta<0$ it flows to the $\cN=2$ sigma-model with target $Gr(N_c,\CC^{|k|})$.Thus for either sign of $k$ we get that the $U(N_c)_k$ Verlinde algebra is isomorphic to the quantum cohomology of $Gr(N_c,\CC^{|k|})$.

We can generalize this argument by adding $N_f$ fundamental flavors ($N_f$ pairs of fundamentals $Q_a$ and anti-fundamentals $\tQ^a$). We give them opposite real masses $m_a,-m_a, a=1,\ldots,N_f$.  The Chern-Simons level in the UV is still $k/2$. Then for $k>0$ and $\zeta<0$ exactly the same argument shows that there is a vacuum with $\sigma=-\zeta/k$ where the low-energy theory is $U(N_c)$ Chern-Simons theory at level $k$ coupled to $N_f$ flavors of fundamentals with real masses $m_a,-m_a$. 
Shifting $\sigma\mapsto \sigma+\zeta/k$ we can set its expectation value to zero and simultaneously set the FI parameter to zero at the expense of shifting the real masses to $m_a-\zeta/k, -(m_a-\zeta/k)$, $a=1,\ldots,N_f$. For $N_f>0$ and depending on the values of $m_a$ this vacuum may be part of the moduli space of vacua where the fields $Q_a$ and $\tQ^a$ get VEVs and Higgs the gauge group. But the fields $q_A$ all have vanishing VEVs because they are all massive. 

On the other hand, for $\zeta>0$ we are forced to set $\sigma=0$, and then the moduli space of the D-flatness equations is the total space of $N_f$ copies of the tautological bundle over $Gr(N_c,\CC^{N_f+k})$.  Let us denote this space $T(N_f,N_c,k)$. Real masses induce a potential on this space  
$$
V=\sum_{a=1}^{N_f} m_a^2\xi_a^2,
$$
where $\xi_a$ is the vector field on $T(k,N_c,N_f)$ corresponding to the $U(1)$ action
$$
Q_a\mapsto \lambda Q_a,\quad \tQ^a\mapsto \lambda^{-1} \tQ^a.
$$

Now if we compactly this theory on a circle, the same arguments as above show that the twisted chiral ring of the 2d $\cN=(2,2)$ sigma-model with target 
$T(N_f,N_c,k)$ deformed by real twisted masses $m_a,-m_a$ must be isomorphic to the Wilson loop algebra in the $U(N_c)_k$ $\cN=2$ Chern-Simons theory coupled to $N_f$ flavors of fundamentals with masses $m_a-c,-(m_a-c)$. We allowed for a possibility that there could be an additional renormalization of the expectation value of $\sigma$ upon compactification to 2d, or equivalently an additional additive renormalization of real masses, and introduced an undetermined constant $c$. It is fixed by the requirement that the 3d gauge theory compactified on a circle develop flat directions for $m_a=c$.

If one wishes, one can deform the theory by a superpotential depending on the fields $Q$ and $\tQ$, this does not affect the argument in any way. In particular, one can choose this superpotential to make the 3d theory $\cN=3$ supersymmetric.

As far as we know, equivariant quantum cohomology of the spaces $T(N_f,N_c,k)$ has not been computed. The GK duality predicts that equivariant quantum cohomology algebras of $T(N_f,N_c,k)$ and $T(N_f,|k|+N_f-N_c,k)$ are isomorphic. Note that these spaces are vector bundles $E^{\oplus N_f}$ and $F^{\oplus N_f}$ over the same Grassmannian $Gr(N_c,\CC^{|k|+N_f})$. Here $E$ is the tautological bundle of rank $N_c$, and $\hat E$ is the complementary bundle of rank $|k|+N_f-N_c$. The bundles in questions fit into an equivariant short exact sequence
$$
0\ra E^{\oplus N_f} \ra V^{\oplus N_f}\ra {\hat E}^{\oplus N_f}\ra 0,
$$
where $V$ is the trivial bundle of rank $|k|+N_f$ with the obvious action of $U(1)^{N_f}$. Hopefully, this exact sequence might enable one to relate the equivariant quantum cohomology algebras of $T(N_f,N_c,k)$ and $T(N_f,|k|+N_f-N_c,k)$ and show that they are indeed isomorphic.

Actually, it appears that a more natural interpretation of our results is in terms of equivariant quantum K-theory of the spaces  $T(N_f,N_c,k)$. Indeed, while the equivariant quantum cohomology of $T(N_f,N_c,k)$ is an algebra over the ring $\ZZ[q,q^{-1},m_1,\ldots,m_{N_f}]$, where the variables $m_a$ live in the Lie algebra of $U(1)^{N_f}$. On the other hand, the equivariant quantum K-theory is an algebra over the representation ring of $U(1)^{N_f}$ tensored with $\ZZ[q,q^{-1}]$, that is, $\ZZ[q,q^{-1},r_1,r_1^{-1},\ldots,r_{N_f},r_{N_f}^{-1}]$, where the variables $r_1,\ldots,r_{N_f}$ live in $U(1)^{N_f}$ or its complexification. In our case the equivariant parameters are the exponentials of the 3d real masses, so it appears that the algebra of Wilson loops should be identified with the equivariant quantum K-theory of the space $T(N_f,N_c,k)$. 

\section{Appendix}

\subsection{Derivation of Expression for $\Phi(-t) \Psi_p(t)$}

Consider the following elements of $\mathcal{R}(U(N))$:

\[ \Phi(t) = \sum_{i=0}^N t^i \phi_i = \prod_{j=1}^N (1 + t x_j) \]

\[ \Psi(t) = \sum_{i=0}^\infty t^i \psi_i = \prod_{j=1}^N (1 - t x_j)^{-1} \]

Now, as in we take a polynomial $p(x)$, and an associated polynomial $\tilde{p}(t)=t^M p(t^{-1})$, which we expand as:

\[ p(x) =  x^M + a_1 x^{M-1} + ... + a_{M-1} x + a_M \]

\[ \tilde{p}(t) = 1 + a_1 t + ... + a_M t^M \]

Now let us define:

\[ \Psi_{p}(t)' = \tilde{p}(t) \Psi(t) = \sum_{i=0}^{\infty} \sum_{j=0}^i a_{j} \psi_{i-j} : = \sum_{i=0}^\infty {\psi_p}_i t^i \]

\[ \Psi_{p}'(t) = \lbrack \Psi_{p}'(t) \rbrack_{M-N} = \sum_{i=0}^{M-N} {\psi_p}_i t^i \]
where we have defined the ${\psi_p}_i$ on the first line as the coefficients of the infinite series $\Psi_p'(t)$, and $\Psi_p(t)$ is the truncation of this series after $M-N$ terms.

We now want to show that, in the quotient ring $\mathcal{A}_p^{(N)}$, we have the relation:

\begin{equation}
\label{mainrel}
\Phi(-t) \Psi_p(t) = \tilde{p}(t) 
\end{equation}
Recall that this quotient is given by $R(U(N))/\mathcal{I}$, where we first define $\mathcal{I}' \subset \mathcal{I}$ by relations:

\[ p(x_j) = 0, \;\; j=1,...,N \]
and then $\mathcal{I}$ is defined by the set of elements $f$ such that $f V \in \mathcal{I}'$, where, in this case, we can take $V$ as the Vandermonde determinant:

\[ V = \prod_{i \neq j} (x_i - x_j) \]
Thus it suffices to show that:

\begin{equation}
\label{cond}
(\Phi(-t) \Psi_p(t)- \tilde{p}(t) ) V \in \mathcal{I}'
\end{equation}

First it is necessary to have a more explicit formula for the $\phi_i$ and $\psi_j$.  This is provided by the Weyl character formula applied to the group $U(N)$, which associates to any partition $\rho=(\rho_1,...,\rho_N)$, with $\rho_1 \geq ... \geq \rho_N \geq 0$, a symmetric function $\theta_\rho(x_1,...,x_N)$ via:

\[ \theta_\rho = \frac{A_{\delta + \rho}}{A_\delta} \]
where $A_\omega$ is defined for a strictly decreasing sequence $\omega=(\omega_1,...,\omega_N)$ with $\omega_1> ... > \omega_N \geq 0$, by:

\[ A_\omega = \sum_{\pi \in S_N} (-1)^\pi \prod_i {x_{\pi(i)}}^{\omega_i} \]
and $\delta = (N-1,N-2,...,0)$, so that $A_\delta=V$. 

Then we have:

\[ \phi_i = \theta_{(\underbrace{\scriptstyle{1,...,1}}_{i},0,...,0)} \]

\[ \psi_j = \theta_{(j,0,...,0)} \]

Thus we can write:

\[ \Psi_p (t) = \sum_{i=0}^{M-N} t^i (\psi_p)_i = \sum_{i=0}^{M-N} t^i \sum_{i=0}^j a_j \psi_{i-j} \]

\[ = \sum_{i=0}^{M-N} t^i \sum_{i=0}^j a_j  \sum_{\pi \in S_N} (-1)^\pi {x_{\pi(1)}}^{N-1+i-j} \prod_k x_{\pi(k)}^{N-k}V^{-1} \]

Now consider the first term in (\ref{cond}):

\begin{equation}
\label{oe}
\Phi(-t) \Psi_p(t) V = \bigg( \prod_{j=1}^N (1 - t x_j) \bigg)\sum_{i=0}^{M-N} t^i \sum_{j=0}^i a_j  \sum_{\pi \in S_N} (-1)^\pi {x_{\pi(1)}}^{N-1+i-j} \prod_{k>1} x_{\pi(k)}^{N-k} 
\end{equation}
where we have canceled $V$ against the denominator in the Weyl character formula applied to $\psi_j$.  Then we get:

\[  \sum_{\pi \in S_N} (-1)^\pi \bigg( \prod_{k>1}(1 - t x_{\pi(k)}) x_{\pi(k)}^{N-k} \bigg)  \sum_{j=0}^{M-N} a_j \sum_{i=j}^{M-N}  t^i {x_{\pi(1)}}^{N-1+i-j} (1 - t x_{\pi(1)})\]
The last sum is a telescoping series, and we're left with:

\[ \sum_{\pi \in S_N} (-1)^\pi \bigg( \prod_{k>1}(1 - t x_{\pi(k)}) x_{\pi(k)}^{N-k} \bigg)\sum_{j=0}^{M-N} a_j  ( t^j {x_{\pi(1)}}^{N-1} - t^{M-N+1} {x_{\pi(1)}}^{M-j} ) \]

\begin{equation}
\label{ttr}
\sum_{\pi \in S_N} (-1)^\pi \bigg( \prod_{k>1}(1 - t x_{\pi(k)}) x_{\pi(k)}^{N-k} \bigg) \bigg( \sum_{j=0}^{M-N} a_j t^j {x_{\pi(1)}}^{N-1} - t^{M-N+1} p(x_{\pi(1)}) + t^{M-N+1} \sum_{j=M-N+1}^M a_j {x_{\pi(1)}}^{M-j}  \bigg) 
\end{equation}
The first and third terms have the form:

\begin{equation}
\label{xtt}
\sum_{\pi \in S_N} (-1)^\pi \bigg( \prod_{k>1}(1 - t x_{\pi(k)}) x_{\pi(k)}^{N-k} \bigg) {x_{\pi(1)}}^{N-n} 
\end{equation}
for some $n \in \{1,...,N\}$.  After expanding the product into a sum of monomials, a given monomial will vanish by antisymmetry unless the powers of the $x_{\pi(k)}$ are distinct.  For $n=1$, this forces us to take the first term in each factor of $(1- t {x_{\pi(k)}})$, and we're left simply with $V$.  If $n>1$, then we see that for $k\leq n$, we must take the second term in $(1- t {x_{\pi(k)}})$, while for $k> n$, we take the first.  Then we see that we obtain $t^n V$, up to a sign which one can check is $1$.  Thus the expression (\ref{xtt}) is simply:

\[ V t^n \]
And we can rewrite the first and third terms in (\ref{ttr}) as:

\[ V \bigg( \sum_{j=0}^{M-N} a_j t^j  + t^{M-N+1} \sum_{j=M-N+1}^M a_j t^{j-M+N-1}  \bigg) = V \tilde{p}(t) \]
Then, reinstating the second term and rearranging, we have the relation:

\begin{equation}
\label{runrel}
( \Phi(-t) \Psi_p(t) - \tilde{p}(t) )V =  - t^{M-N+1} \sum_{\pi \in S_N} (-1)^\pi  \prod_{k>1}(1 - t x_{\pi(k)}) x_{\pi(k)}^{N-k} p(x_{\pi(1)}) 
\end{equation}
This relation is valid in $R(U(N))$, or any quotient thereof.   We can see that the RHS is in $\mathcal{I}'$, and so we obtain (\ref{cond}) as desired.

If we denote by ${\mathcal{A}'}_p^{(N)}$ the ring generated by the symbols $\phi_i$ and ${\psi_p}_j$ subject to the relation (\ref{mainrel}), then what we have just shown amounts to the statement that there exists a homomorphism:

\[ g:{\mathcal{A}'}_p^{(N)}\rightarrow \mathcal{A}_p^{(N)} \]
sending $\phi_i$ and ${\psi_p}_j$ to the (equivalence class containing) the symmetric polynomial defined above.  Although we will not prove it here, we conjecture that this is actually an isomorphism.  

\subsection{Derivation of Inductive Formula for $(N,k,N_f)$ Wilson Loop Expectation Values}

In this section we prove the formula (\ref{formula}) expressing the Wilson loop expectation value in $U(N)$ theories at level $k$ with $N_f$ flavors inductively in terms of those in theories with lower $N_f$. We follow the notation of Section \ref{sec:wloopmap}.

The formula will follow from considering two ways of rewriting the following quantity:

\[ I = \left< \Phi(r_b) \Psi_{p_b}(-r_b) \prod_\alpha \Phi(-t_\alpha) \prod_\beta\Psi_p(t_\beta) \right>_{N,k,N_f;\zeta,m_a} \]
First, note that $\Phi(r_b) = \prod_j (1 + e^{2 \pi (\lambda_j + m_b)})$, and this cancels partially against the contribution of the $b$th flavor in the matrix model.  There are extra factors of $e^{\pi \lambda_j}$ which can be absorbed into a shift of $\zeta$, and we find:

\[ I = e^{2 \pi i \alpha_1} \left< \Psi_{p_b}(-r_b) \prod_\alpha \Phi(-t_\alpha) \prod_\beta\Psi_p(t_\beta)\right>_{N,k,N_f-1;\zeta-\frac{i}{2},m_a/m_b} \]
where we have defined:

\[ \alpha_1(N,k,N_f;\zeta,m_a;m_b) = \delta(N,k,N_f;\zeta,m_a)-\delta(N,k,N_f-1;\zeta-\frac{i}{2},m_a/m_b) -\frac{i}{2} N m_b \]
accounting also for the new phase canceling factor we must insert since $N_f$ has changed.

It is convenient to rewrite this insertion of $\Psi_p$ in terms of $\Psi_{p_b}$.  Note that, even in the $(N_f-1)$ -flavor algebra, we have:

\[ \Phi(-t) \Psi_p(t) = \tp(t) \]
This is because this relation uses only the vanishing of $p(x)$, which also holds in the algebra $\mathcal{A}_{p_b}^{(N)}$, since $p_b(x)$ divides $p(x)$.  But note that this is precisely the same relation satisfied by the quantity $\Psi_{p_b}(t) (1+t {r_b}^{-1})$.  Since these relations can be used to solve for the coefficients of $\Psi_p(t)$ and $\Psi_{p_b}(t) (1+t r_b)$ in terms of the $\phi_i$, they must be equal (in the $(N_f-1)$ flavor theory):

\[ \Psi_p(t) = (1 + t {r_b}^{-1}) \Psi_p(t) \]
Thus we can write:

\[ I = e^{2 \pi i \alpha_1} \left< \Psi_{p_b}(-r_b) \prod_\alpha \Phi(-t_\alpha) \prod_\beta (1 + r_b t_\beta)\Psi_{p_b}(t_\beta)\right>_{N,k,N_f-1;\zeta-\frac{i}{2},m_a/m_b} \]

Now let us write $I$ a different way.  We can use (\ref{runrel}) to rewrite the product $\Phi(r_b) \Psi_{p_b}(-r_b)$, and we get:

\[ I = \left< \bigg( \tp_b(-r_b) - (-r_b)^{k+N_f-N+1} V^{-1} \sum_{\pi \in S_N} (-1)^\pi  \prod_{j>1}(1 + r_b x_{\pi(j)}) x_{\pi(j)}^{N-j} p_b(x_{\pi(1)})  \bigg)\prod_\alpha \Phi(-t_\alpha) \prod_\beta\Psi_p(t_\beta)\right>_{N,k,N_f;\zeta,m_a} \]

\[ = \tp_b(-r_b) \left< \prod_\alpha \Phi(-t_\alpha) \prod_\beta\Psi_p(t_\beta) \right>_{N,k,N_f;\zeta,m_a} - \]

\[ -  N! (-r_b)^{k+N_f-N-1} \left< V^{-1} p_b(x_1) \prod_{j>1} (1 + r_b x_j) {x_j}^{N-j} \prod_\alpha \Phi(-t_\alpha) \prod_\beta\Psi_p(t_\beta) \right>_{N,k,N_f;\zeta,m_a} \]
Note that the second term is not zero because $p_b(x_1)$ does not vanish in the $(N,k,N_f)$ theory.  When we attempt to repeat the contour shifting argument, we find that the poles are no longer all canceled by $p_b(x)$, specifically, there are poles from the $b$th flavor at $\lambda_1 = -m_b + \frac{i}{2}$.  To see this explicitly, we write (using a mixed notation with both $\lambda_j$ and $x_j = e^{2 \pi \lambda_j}$):

\[ N! \left< V^{-1} p_b(x_1) \prod_{j>1} (1 + r_b x_j) {x_j}^{N-j} \prod_\alpha \Phi(-t_\alpha) \prod_\beta\Psi_p(t_\beta) \right>_{N,k,N_f;\zeta,m_a} = \]

\[ = \int d^N \lambda \prod_j \frac{ e^{-k \pi i {\lambda_j}^2 + 2 \pi i \zeta \lambda_j}}{\prod_a 2 \cosh \pi (\lambda_j + m_a)} \prod_{i \neq j } 2 \sinh \pi (\lambda_i - \lambda_j)  \frac{ p_b(x_1) \prod_{j>1} (1 + r_b x_j) {x_j}^{N-j}}{\prod_{i<j} (x_i - x_j)} \prod_\alpha \Phi(-t_\alpha) \prod_\beta\Psi_p(t_\beta) \]

\[ = \int d^N \lambda \prod_j \frac{ e^{-k \pi i {\lambda_j}^2 + 2 \pi i \zeta \lambda_j}}{\prod_a 2 \cosh \pi (\lambda_j + m_a)} \prod_{i > j } 2 \sinh \pi (\lambda_i - \lambda_j) {x_1}^{1-N} p_b(x_1) \prod_{j>1} (1 + r_b x_j) {x_j}^{(N-j)/2}\prod_\alpha \Phi(-t_\alpha) \prod_\beta\Psi_p(t_\beta) \]
Now consider the integral over $\lambda_1$, plugging in the form of $p_b(x_1)$ from above:

\[ \int d\lambda_1 \frac{ e^{-k \pi i {\lambda_1}^2 + 2 \pi i \zeta \lambda_1}}{\prod_a 2 \cosh \pi (\lambda_1 + m_a)} \prod_{i > 1 } 2 \sinh \pi (\lambda_i - \lambda_1) {x_1}^{1-N} (1 - (-1)^{k+N_f} q^{-1} {x_1}^k) \prod_{a \neq b} ( x_1 r_a + 1) \prod_\alpha \Phi(-t_\alpha) \prod_\beta\Psi_p(t_\beta) \]

By the usual logic, the factor of $1 - (-1)^{k+N_f} q^{-1} {x_1}^k$ allows us to rewrite this integral as a contour integral along the boundary $\mathcal{C}$ of the region $0< \mbox{Im} \lambda_1 < 1$:

\[ = \int_{\mathcal{C}} \frac{ e^{-k \pi i {\lambda_1}^2 + 2 \pi i \zeta \lambda_1}}{\prod_a 2 \cosh \pi (\lambda_1 + m_a)} \prod_{i > 1 } 2 \sinh \pi (\lambda_i - \lambda_1)e^{2 \pi (1-N) \lambda_1} \prod_{a \neq b} ( e^{2 \pi (\lambda_1 + m_a)} + 1) \prod_\alpha \Phi(-t_\alpha) \prod_\beta\Psi_p(t_\beta) \]
We see all the poles in the denominator are canceled except the one at $\lambda_1 = -m_b + \frac{i}{2}$.  Evaluating the residue here gives:

\[ \exp \bigg( - k \pi i (-m_b + \frac{i}{2})^2 + 2 \pi i \zeta (-m_b + \frac{i}{2}) + 2 \pi m_b ( N - \frac{N_f+1}{2} ) + \frac{\pi i}{2} (N_f-N) + \pi \sum_{a \neq b} m_a \bigg) \times \]

\[ \times \prod_{j>1} 2 \cosh \pi (\lambda_j + m_b)  \bigg( \prod_\alpha \Phi(-t_\alpha)\prod_\beta\Psi_p(t_\beta) \bigg) \bigg|_{x_1 = - {r_b}^{-1}}\]
Plugging this back into the expression above, we find the contribution of the $b$th flavor is canceled, and we are left with (denoting by $e^{i \delta}$ the first line in the last equation):

\[ e^{i \delta} \int d^{N-1} \lambda \prod_{j>1} \frac{ e^{-k \pi i {\lambda_j}^2 + 2 \pi i \zeta \lambda_j+ \pi (N-j) \lambda_j}}{\prod_{a \neq b} 2 \cosh \pi (\lambda_j + m_a)} \prod_{i > j } 2 \sinh \pi (\lambda_i - \lambda_j) \prod_{j>1} (1 + r_b x_j) \bigg( \prod_\alpha \Phi(-t_\alpha)\prod_\beta\Psi_p(t_\beta) \bigg) \bigg|_{x_1 = - {r_b}^{-1}}\]

\[ = e^{2 \pi i \alpha_2} \left< \Phi(r_b) \bigg( \prod_\alpha \Phi(-t_\alpha) \prod_\beta\Psi_p(t_\beta) \bigg) \bigg|_{x_1=-{r_b}^{-1}} \right>_{N-1,k,N_f-1;\zeta+\frac{i}{2},m_a/m_b} \]
where we have defined:

\[ \alpha_2(N,k,N_f;\zeta,m_a;m_b) = \delta(N,k,N_f;\zeta,m_a) -\delta(N-1,k,N_f-1;\zeta+\frac{i}{2},m_a/m_b) - \]

\[ - \frac{i}{2} (k+N_f-N) m_b  + \frac{i}{2} \sum_{a \neq b} m_b - \frac{1}{2} k {m_b}^2 - \frac{i}{2} \zeta - \zeta m_b + \frac{1}{8} k  + \frac{1}{4} N_f - \frac{1}{4} \]

Consider the factor:

\[ \bigg( \prod_\alpha \Phi(-t_\alpha) \prod_\beta\Psi_p(t_\beta) \bigg) \bigg|_{x_1=-{r_b}^{-1}} \]
When we plug in $x_1=-{r_b}^{-1}$ to $\Phi(-t_\alpha)$, we get simply:

\[ (1 + {r_b}^{-1} t_\alpha) \prod_{j>1} (1 - t_\alpha x_j) \rightarrow (1 + {r_b}^{-1}t_\alpha) \Phi(-t_\alpha) \]
where on the RHS we mean $\Phi(-t_\alpha)$ in the $(N-1)$-color theory.  Similarly, for $\Psi(t_\beta)$, it is easiest to plug this in to the relation (\ref{runrel}):

\[ ( \Phi(-t_\beta) \Psi_p(t_\beta) - \tilde{p}(t) )\bigg|_{x_1=-{r_b}^{-1} }= =  - \bigg( V^{-1} t^{M-N+1} \sum_{\pi \in S_N} (-1)^\pi  \prod_{k>1}(1 - t_\beta x_{\pi(k)}) x_{\pi(k)}^{N-k} p(x_{\pi(1)}) \bigg) \bigg|_{x_1 = -{r_b}^{-1}} \]
But the RHS vanishes, because $p(x_j)=0$ for $j>1$ in the $(N-1)$-color theory and $p(-{r_b}^{-1})=0$.  Thus we have:

\[  (1 + {r_b}^{-1} t_\beta)  \Phi(-t_\beta) \bigg( \Psi_p(t_\beta) \bigg) \bigg|_{x_1=-{r_b}^{-1}} = \tp(t_\beta) \]

\[ \Rightarrow \Phi(-t_\beta) \bigg( \Psi_p(t_\beta) \bigg) \bigg|_{x_1=-{r_b}^{-1}} = \tp_b(t_\beta) \]
But this is precisely the same relation satisfied by $\Psi_{p_b}(t_\beta)$.  By a similar argument as before, this means that we must have, in the $(N-1,k,N_f-1)$ theory:

\[ \bigg( \Psi_p(t_\beta) \bigg) \bigg|_{x_1=-{r_b}^{-1}} = \Psi_{p_b}(t_\beta) \]
Putting this together, we find the second expression for $I$:

\[ I = \tp_b(-r_b) \left< \prod_\alpha \Phi(-t_\alpha) \prod_\beta\Psi_p(t_\beta) \right>_{N,k,N_f;\zeta,m_a} - e^{2 \pi i \alpha_2} \left< \Phi(r_b) \prod_\alpha (1+{r_b}^{-1} t_\alpha) \Phi(-t_\alpha) \prod_\beta\Psi_{p_b}(t_\beta) \right>_{N-1,k,N_f-1;\zeta+\frac{i}{2} ,m_a/m_b} \]

Equating these two expressions for $I$ and rearranging to solve for the $N_f$-flavor expectation value, we find:

\[  \left< \prod_\alpha \Phi(-t_\alpha) \prod_\beta\Psi_p(t_\beta) \right>_{N,k,N_f;\zeta,m_a} = \]

\[ = \frac{1}{\tp_b(-r_b)} \bigg( e^{2 \pi i \alpha_1}  \left< \Psi_{p_b}(-r_b) \prod_\alpha \Phi(-t_\alpha) \prod_\beta(1 + t_\beta {r_b}^{-1}) \Psi_{p_b} (t_\beta) \right>_{N,k,N_f-1;\zeta-\frac{i}{2},m_a/m_b}  + \]

\[ + e^{2 \pi i \alpha_2} \left< \Phi(r_b) \prod_\alpha (1 + t_\alpha {r_b}^{-1}) \Phi(-t_\alpha) \prod_\beta \Psi_{p_b} (t_\beta)  \right>_{N-1,k,N_f-1;\zeta+\frac{i}{2},m_a/m_b} \bigg) \]
agreeing with (\ref{formula}).

The final step is to prove the relation (\ref{phaserel}) between $\alpha_1$ and $\alpha_2$:

\[ \alpha_2(N,k,N_f;\zeta,m_a;m_b) = \alpha_1(k+N_f-N,-k,N_f;-\zeta,m_a;m_b) \]

From above, we have:

\[ \alpha_2(N,k,N_f;\zeta,m_a;m_b) = \delta(N,k,N_f;\zeta,m_a) -\delta(N-1,k,N_f-1;\zeta+\frac{i}{2},m_a/m_b) - \]

\[ - \frac{i}{2} (k+N_f-N) m_b + \frac{i}{2} \sum_{a \neq b} m_b - \frac{1}{2} k {m_b}^2 - \frac{i}{2} \zeta - \zeta m_b + \frac{1}{8} k  + \frac{1}{4} N_f - \frac{1}{4} \]
and:

\[ \alpha_1(k+N_f-N,-k,N_f;-\zeta,m_a;m_b) = - \alpha_1(k+N_f-N,k,N_f;\zeta^*,{m_a}^*;{m_b}^*)^* \]

\[ = - ( \delta(k+N_f-N,k,N_f;\zeta^*,{m_a}^*)-\delta(k+N_f-N,k,N_f-1;\zeta^*-\frac{i}{2},{m_a}^*/{m_b}^*) -\frac{i}{2} (k+N_f-N) {m_b}^* )^* \]

\[ = \delta(k+N_f-N,-k,N_f;-\zeta,m_a) - \delta(k+N_f-N,-k,N_f-1;-\zeta-\frac{i}{2},m_a/m_b)- \frac{i}{2} (k+N_f-N) m_b \]
Subtracting these two quantities, we find:

\[ \delta(N,k,N_f;\zeta,m_a) -\delta(N-1,k,N_f-1;\zeta+\frac{i}{2},m_a/m_b)+ \]

\[- \frac{i}{2} (k+N_f-N) m_b + i m_b + \frac{i}{2} \sum_{a \neq b} m_b - \frac{1}{2} k {m_b}^2 + \frac{i}{2} \zeta - \zeta m_b + \frac{1}{8} k  + \frac{1}{4} N_f + \frac{1}{4} -  \]

\[ - \delta(k+N_f-N,-k,N_f;-\zeta,m_a) + \delta(k+N_f-N,-k,N_f-1;-\zeta-\frac{i}{2},m_a/m_b) + \frac{i}{2} (k+N_f-N) m_b  \]
Rewriting this in terms of $\gamma(k,N_f;\zeta,m_a)=\delta(k+N_f-N,-k,N_f;-\zeta,m_a)-\delta(N,k,N_f;\zeta,m_a)$ and simplifying, we get:

\[ -\gamma(k,N_f;\zeta,m_a) +\gamma(k,N_f-1;\zeta+\frac{i}{2},m_a/m_b)+ \]

\[ + \frac{i}{2} \sum_{a \neq b} m_b - \frac{1}{2} k {m_b}^2 - \frac{i}{2} \zeta - \zeta m_b + \frac{1}{8} k + \frac{1}{4} N_f - \frac{1}{4} \]
Finally, we need like to check that the definition for $\gamma$ given in (\ref{phase}) solves this inductive relation.  Plugging in that definition, we find that the first two terms give:

\[ -( \frac{1}{24} (k^2 + 3 (k+N_f) (N_f-2) + 2) + \frac{1}{2} \zeta^2 - \frac{1}{2} k \sum_a {m_a}^2 - \zeta \sum_a m_a) +\]

\[ +  (\frac{1}{24} (k^2 + 3 (k+N_f-1) (N_f-3) + 2) + \frac{1}{2} (\zeta+\frac{i}{2})^2 - \frac{1}{2} k \sum_{a \neq b} {m_a}^2 - (\zeta + \frac{i}{2}) \sum_{a \neq b} m_a ) \]

\[ = -\frac{1}{4} N_f - \frac{k}{8} + \frac{1}{4} + \zeta(\frac{i}{2} + m_b) + \frac{1}{2} k {m_b}^2 - \frac{i}{2} \sum_{a \neq b} m_a \]
which cancels precisely against the remaining terms.  This completes the proof.

\bibliographystyle{JHEP}
\bibliography{wloopdraftfinal}

\end{document}